\documentclass[prd,aps,preprint,preprintnumbers,nofootinbib]{revtex4}
\usepackage{epsfig,footnote}
\usepackage{ulem}
\usepackage{color}
\usepackage{array}
\usepackage{amssymb}
\usepackage{amsmath}
\usepackage{graphicx,subfigure}
\usepackage{longtable}
\usepackage{verbatim}
\usepackage{amsfonts}
\usepackage{hyperref}
\usepackage{cancel}

\begin{document}
\title{Analysis of three-body charmless $B$-meson decays under the factorization-assisted
topological-amplitude approach}
\author{Si-Hong Zhou\footnote{shzhou@imu.edu.cn}, Xin-Xia Hai and Run-Hui Li\footnote{lirh@imu.edu.cn} }
\affiliation{School of Physical Science and Technology,
Inner Mongolia University, Hohhot 010021, China}
\author{Cai-Dian L\"u\footnote{lucd@ihep.ac.cn}}
\affiliation{Institute of High Energy Physics, CAS, Beijing 100049, China }
\affiliation{ School of Physics, University of Chinese Academy of Sciences, Beijing 100049, China
}
\date{\today}
\begin{abstract}
We analyze quasi-two-body charmless $B$ decays $B_{(s)} \to P_1 V  \to  P_1 P_2 P_3$ 
with $V$ representing a vector resonant, and $P_{1,2,3}$ as a light 
pseudoscalar meson, pion, kaon or $\eta^{(\prime)}$. The intermediate processes $B_{(s)} \to P_1 V $ 
are calculated in the factorization-assisted topological-amplitude approach and the vector resonant 
effects are described by the Breit-Wigner propagator, which successively decay to $P_1 P_2$ via 
strong interaction. Taking into account of all vector resonances in ground state, $\rho, K^*, \omega, \phi$, 
we present the related branching fractions, and calculate the virtual effects for 
$B_{(s)} \to \pi, K (\rho ,\omega \to) KK$. 
We also predict direct $\it{CP}$ asymmetries of three-body $B$ decay modes with  $\rho, K^*$ resonances as 
intermediate states. Our predicted
branching fractions of decay modes dominated by the color-favored tree diagram or the color-favored 
penguin diagram are consistent with the perturbative QCD (PQCD) approach's predictions as well as 
QCD factorization approach. While for those nonperturbative contribution dominated decay modes,
the branching ratios in this work are   in better agreement with current experimental data than the PQCD predictions and the   
QCD factorization results due to their shortage of the nonperturbative contributions or $1/m_b$ power 
corrections. Many of the decays channels, especially for direct $\it{CP}$ 
asymmetries,  are waiting for  the future experiments.
\end{abstract}

\maketitle

\section{Introduction}\label{Introduction}

Three-body nonleptonic $B$ meson decays constitute a large portion of $B$ decay channels.
Contrary to two-body $B$ meson decays, they have nontrivial kinematics and 
phase space distributions, which can provide more opportunities for the study of 
${CP}$ asymmetries and factorization issues in QCD for   multibody 
nonleptonic decays. Generally, one utilize the Dalitz plot technique to analyze the phase space of
three-body $B$ meson decays, where the invariant mass of a pair of final-state particles peaks as a resonance.
In the edges of the Dalitz plot, various resonances as the intermediate states in 
three-body $B$ meson decays will show up,  so that the analysis on these decays 
enables one to study the properties of these resonances. 

Experimentally, a large amount of  branching ratios and ${CP}$ asymmetry parameters of 
three-body charmless $B$ decays have already been measured by BABAR, Belle and LHCb 
collaborations~\cite{BaBar:2009vfr, Nasteva:2013bta, LHCb:2013lcl, LHCb:2014mir, BaBar:2015pwa}, 
while more are expected from the upgrade of LHCb and Belle II experiments.
On the theoretical side, three-body nonleptonic $B$ meson decays are more complicated 
than two-body cases as they involve three-body decay matrix elements.  In a phenomenological 
factorization model~\cite{Cheng:2002qu, Cheng:2007si, Cheng:2013dua},
 the three-body matrix elements of charmless $B$ decays were factorized naively, with 
 the resulting local correlaters studied extensively based on heavy meson chiral 
 perturbation theory  for nonresonant contributions and the usual 
 Breit-Wigner formalism for resonant effects. 
 Besides these phenomenological descriptions of three-body charmless $B$ decays in the 
 whole Dalitz plot,   theoretical analysis in the context of QCD factorization   mainly 
 concentrated on the edges of the Dalitz distribution, since the contribution in the center 
 regions of phase space is argued to be both $1/m_b$ power and $\alpha_s$ 
 suppressed~\cite{Krankl:2015fha,Chen:2002th}.
 
  Since the vector (scalar) mesons usually decay dominantly to two pseudoscalar mesons with large decay width, the three-body $B$ decays are found to be dominantly happened by $B$ decaying to a vector  (scalar) meson and a pseudoscalar meson, and the vector  (scalar) meson subsequently decaying to two pseudoscalar mesons.
In this case, two of the three final particles are collinear, generating a small 
invariant mass in the edges of the Dalitz distribution, and recoil against the third meson 
in three-body decays. Then one expects a similar 
factorization theorem as for two-body decays, with the difference that one of the mesons 
in the two-body case is substituted by a pair of collinear moving mesons with small 
invariant mass. A series of works~\cite{Chen:2002th,Wang:2014ira,Wang:2016rlo, 
Li:2016tpn, Li:2018psm, Wang:2020plx, Fan:2020gvr, Wang:2020nel, Zou:2020atb, Zou:2020fax,Zou:2020ool,Yang:2021zcx,Liu:2021sdw} using perturbative 
QCD approach concentrated on this situation that the pair of collinear mesons are 
described collectively through intermediate resonances, that is, the three-body decays 
are processed as quasi-two-body decays.
The $\it{CP}$ asymmetry of $B \to \pi\pi\pi$ decays has been studied within the 
QCD factorization framework at leading order in $\alpha_s$~\cite{Klein:2017xti, Mannel:2020abt}, using naive 
factorization for three-body matrix elements, which involve generalized form factor of $B \to \pi \pi$ transition
and dimeson light cone distribution amplitudes in~\cite{Cheng:2017smj, Descotes-Genon:2019bud, Hambrock:2015aor, 
Cheng:2017sfk}. The calculation of the nonleptonic three-body $B$ decays with heavy-to-heavy 
transitions, up to the precision of next-next-to-leading order in $\alpha_s$ 
has also been done in QCD factorization  \cite{Huber:2020pqb}.
 In all these quasi-two-body calculations of the three-body $B$ decays, 
 the theoretical calculations rely heavily on the
 precision of two-body theoretical calculation methods.
 
In the case of two-body nonleptonic $B$ decays, power corrections in the heavy-quark 
expansion, such as weak annihilation 
effects and chirally enhanced power corrections, are difficult to estimate and remain the 
sources of uncertainty in QCD factorization approach~\cite{Cheng:2009cn}. A complete next-to-leading order calculation of two-body $B$ decays in perturbative QCD factorization approach has not been finished \cite{Chai:2022ptk}.
The problem about precision predictions in the 
calculation of two-body nonperturbative matrix elements  also exist in three-body case.
 As discussed in Ref~\cite{Cheng:2013dua}, the predicted branch ratios of quasi-two-body 
 decays such as $B \to \phi (K^+ K^-) K, K^* (K^- \pi^+)\pi, \rho (\pi^+ \pi^-)K$ are smaller 
 than experimental data due to the absence of $1/m_b$ power corrections.
To include the nonperturbative and nonfactorization contribution in two-body $B$ decays, a model independent framework of topological diagram approach
\cite{Cheng:2014rfa} has been introduced.   The equivalence of the topological decay amplitudes and the $SU(3)$ irreducible amplitudes has been approved \cite{He:2018joe,He:2018php}. In this approach, one classifies the decay amplitudes of two-body charmless $B$ decays into different electroweak topological
Feynman diagrams under SU(3) symmetry. The topological amplitudes including the nonfactorizable QCD 
contributions  were extracted through a global fit from 
all experimental data of these decays. The precision of this topological diagram approach is limited to the size of SU(3) breaking effect.

In order to include  the SU(3) breaking effects, the so-called factorization-assisted topological-amplitude 
(FAT) approach~\cite{Li:2012cfa, Li:2013xsa, Zhou:2015jba, Zhou:2016jkv, Jiang:2017zwr,
Zhou:2019crd, Zhou:2021yys,Qin:2021tve} has been introduced. They can give the most precise decay amplitudes of the two-body charmless $B$-meson decays. For instance, the amplitude of the color-suppressed 
topological diagram dominated by the nonfactorizable QCD effect, was larger in the FAT approach 
than that in other perturbative approaches so that the long-standing $B \to \pi \pi$ branching ratio puzzle and 
$B \to \pi K$ $\it{CP}$ asymmetry puzzle were  solved simultaneously.
Encouraged by the success in two-body $B$ decays, we will systematically analyze three-body charmless $B$ decays using the FAT approach
at the edges of the Dalitz distribution with only 
vector resonance contribution, i.e., quasi-two-body decays 
with two of the pseudoscalar mesons decaying from the vector resonance. 
The vector resonant effects are described in terms of the usual 
Breit-Wigner formalism, and a strong coupling accounts for the subsequent vector meson two-body decay. 
We will study the branching ratios and $\it{CP}$ violations 
of these three-body $B$ decays, and discuss the virtual effects of intermediate 
resonance $\rho, K^*, \omega$ , and  $\phi$  on quasi-two-body decays.

This paper is organized as follows. In Sec. \ref{sec:2}, the theoretical framework is introduced. 
The numerical results and discussions are collected in Sec. \ref{sec:3}. 
Section \ref{sec:4} is a summary.


\section{Factorization Amplitudes for Topological Diagrams}\label{sec:2}

\begin{figure} [htb]
\begin{center}
\scalebox{1}{\epsfig{file=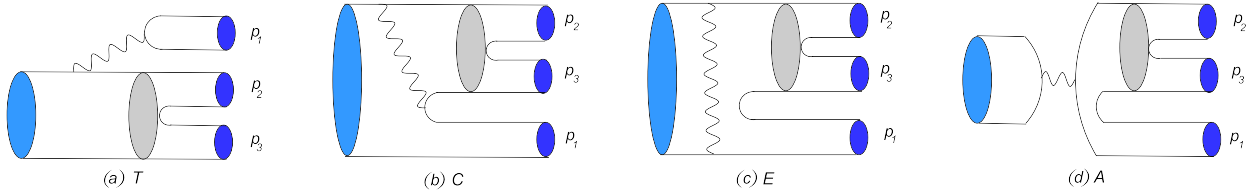}}
\caption{Typical topological tree diagrams of $B_{(s)} \to P_1 V  \to  P_1 P_2 P_3$ under the 
framework of quasi-two-body decay with the wavy line representing a $W$ boson:
  (a) color-favored tree diagram $T$,
  (b) color-suppressed tree diagram $C$, 
  (c) $W$-exchange tree diagram $E$,
  and (d) $W$-annihilation tree diagram $A$.}
\label{tree}
\end{center}
\end{figure}

\begin{figure}[htb]
\begin{center}
\scalebox{1}{\epsfig{file=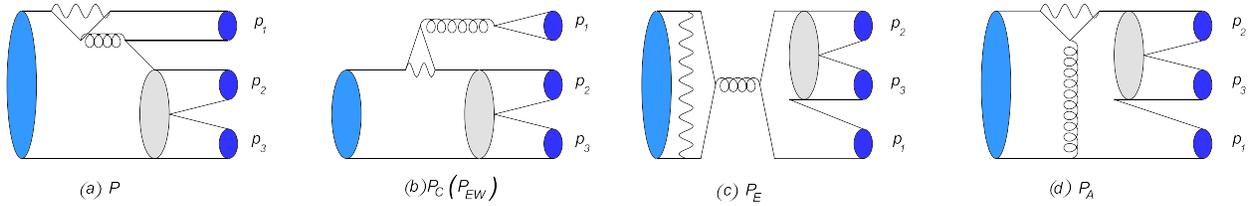}}
\caption{Typical topological penguin diagrams of $B_{(s)} \to P_1 V  \to  P_1 P_2 P_3$ under the 
framework of quasi-two-body decay with the wavy line representing a $W$ boson:
   (a) QCD-penguin emission diagram $P$,
   (b) flavor-singlet QCD-penguin diagram $P_C$
        or EW-penguin diagram $P_\mathrm{EW}$,
   (c) timelike QCD-penguin diagram $P_E$ and
   (d) spacelike QCD-penguin annihilation diagram $P_A$.}
\label{penguin}
\end{center}
\end{figure}

As  a quasi-two-body process, the $B_{(s)}  \to  P_1 P_2 P_3$ decay (with $P$ denoting  a light 
pseudoscalar meson) is 
divided into two stages. First, $B_{(s)}$ meson decays to $P_1 V$,  and the intermediate 
vector resonance $V$ decays to $ P_2 P_3$ subsequently. 
The first decay $B_{(s)} \to P_1 V$ is a weak decay induced by $b \rightarrow \,u\, \, \bar{u} \, d(s)\, $ 
at quark level in leading order (tree diagram) and $b \rightarrow \,d (s)\, \, q\, \bar{q} \, \, (q=u, d, s)$ 
in next-to-leading order (penguin loop diagram). The secondary decay $V \to  P_2 P_3$ proceeds 
via strong interaction. In Fig.~\ref{tree} we show the topological diagrams of 
$B_{(s)}  \to  P_1 P_2 P_3$ under the framework of quasi-two-body decay at tree level, 
including (i) color-favored tree emission diagram $T$, (ii) color-suppressed tree emission diagram $C$,  
(iii) $W$-exchange tree diagram $E$, and (iv) $W$-annihilation tree diagram $A$, which are specified 
by topological structures of the weak interaction. Likewise, one loop penguin diagram can also be 
grouped into four categories: (i) QCD-penguin emission diagram $P$,
(ii) flavor-singlet QCD-penguin diagram $P_C$ or EW-penguin emission diagram $P_{EW}$,
(iii) time-like QCD-penguin diagram $P_E$, and (iv) space-like QCD-penguin annihilation diagram $P_A$, 
which are shown in Fig.~\ref{penguin}. In the two kinds of figures, we only show one case that 
the intermediate resonance (labeled by gray ovals) is produced as a recoiling particle in emission diagrams.
 
The electroweak $B$ decays at tree level have been proven to be factorizable
at high precision \cite{Bauer:2000yr}. 
Therefore, we can calculate the decay amplitude of color-favored tree diagram $T$ in perturbative QCD order by order. Large nonfactorizable contributions, 
such as soft corrections to the color-suppressed tree amplitude and $1/m_b$ power 
corrections from annihilation type diagram, are found to be non-negligible ~\cite{Zhou:2016jkv}. Similar to the FAT approach for two-body $B$ decays~\cite{Zhou:2016jkv}, we have to introduce 
unknown parameters for nonfactorizable contributions for color-suppressed tree diagram $C$, and
 $W$-exchange tree diagram $E$ ($W$-annihilation tree diagram$A$ is negligible) to be fitted from experimental data.
 The formulas of $B_{(s)} \to  P_1 V,\, V P_1$  decays
are given as ~\cite{Zhou:2016jkv}
\begin{align}
\begin{aligned}
T^{P_1V}&=\sqrt{2}\, G_{F}\, V_{ub}V^*_{uq^{'}}\, a_{1}
(\mu)\, f_{V}\,  m_{V}\, F_{1}^{B-P_1}(m_{V}^{2})\, (\varepsilon^{*}_{V}\cdot p_{B}),\\
T^{VP_1}&=\sqrt{2}\, G_{F}\, V_{ub}V^*_{uq^{'}}\, a_{1}
(\mu)\, f_{P_1}\,  m_{V}\, A_{0}^{B-V}(m_{P_1}^{2})\, (\varepsilon^{*}_{V}\cdot p_{B}),
\\
C^{P_1V}&=\sqrt{2}\, G_{F}\, V_{ub}V^*_{uq^{'}}\, \chi^{C^{\prime}}\mathrm{e}^{i\phi^{C^{\prime}}}\, 
         f_{V} \, m_{V}\, F_{1}^{B-P_1}(m_{V}^{2})\, (\varepsilon^{*}_{V}\cdot p_{B}),
\\
C^{VP_1}&=\sqrt{2}\, G_{F}\, V_{ub}V^*_{uq^{'}}\, \chi^{C}\mathrm{e}^{i\phi^{C}}\, 
        f_{P_1}\,  m_{V}\, A_{0}^{B-V}(m_{P_1}^{2})\, (\varepsilon^{*}_{V}\cdot p_{B}),\\
E^{P_1V,VP_1} &=\sqrt{2}\, G_{F}\, V_{ub}V^*_{uq^{'}}\, \chi^{E} \mathrm{e}^{i\phi^{E}}\, 
f_{B}\, m_{V}\,\left (\frac{f_{P_1}f_{V}}{f_{\pi}^{2}}\right)\, (\varepsilon^{*}_{V}\cdot p_{B}).
\end{aligned}
\end{align}
The QCD-penguin emission diagram $P$ and the electroweak penguin emission diagram $P_\mathrm{EW}$ are also proved to be factorizable, and thus are calculable in QCD factorization. 
The flavor-singlet QCD-penguin diagram $P_C$  and spacelike QCD-penguin annihilation diagram $P_A$ 
contain large nonfactorizable contribution which will be determined by $\chi^2$ fit from experimental data. 
The timelike QCD-penguin diagram $P_E$ is found to be smaller than other contribution, 
which can be ignored in the modes not dominated by it in the fit as discussed in~\cite{Zhou:2016jkv}. 
The formulas can be expressed as~\cite{Zhou:2016jkv}
\begin{align}
\begin{aligned}
P^{P_1V}&=-\sqrt{2}\, G_{F}\,  V_{tb}V_{tq^{'}}^{*}\, a_{4}(\mu)\,  f_{V}\, m_{V}\, F_{1}^{B-P_1}(m_{V}^{2})\, 
(\varepsilon^{*}_{V}\cdot p_{B}),\\
P^{VP_1}&=-\sqrt{2}\, G_{F}\, V_{tb}V_{tq^{'}}^{*}\, \left[a_{4}(\mu)-\chi^{P} \mathrm{e}^{i\phi^{P}}r_{\chi}\right]\, 
 f_{P_1}\, m_{V}\, A_{0}^{B-V}(m_{P_1}^{2})\, (\varepsilon^{*}_{V}\cdot p_{B}),\\
 P_{C}^{P_1V}&=-\sqrt{2}\, G_{F}\, V_{tb}V_{tq^{'}}^{*} \, \chi^{P_C^{\prime}}\mathrm{e}^{i\phi^{P_C^{\prime}}}\, 
f_{V}\, m_{V}\, F_{1}^{B-P_1}(m_{V}^{2})\, (\varepsilon^{*}_{V}\cdot p_{B}),\\
P_{C}^{VP_1}&=-\sqrt{2}\, G_{F}\, V_{tb}V_{tq^{'}}^{*}\,  \chi^{P_C}\mathrm{e}^{i\phi^{P_C}}\, 
f_{P_1}\, m_{V}\, A_{0}^{B-V}(m_{P_1}^{2})\, (\varepsilon^{*}_{V}\cdot p_{B}),\\
P_{A}^{P_1V}&=-\sqrt{2}\, G_{F}\, V_{tb}V_{tq^{'}}^{*}\, \chi^{P_{A}}\mathrm{e}^{i\phi^{P_{A}}}\, 
f_{B}\, m_{V}\, \left(\frac{f_{P_1}f_{V}}{f_{\pi}^{2}}\right)\, (\varepsilon^{*}_{V}\cdot p_{B}),\\
\end{aligned}
\end{align}
\begin{align}
\begin{aligned}
P_{EW}^{P_1V}&=-\frac{3\sqrt{2}}{2}\, G_{F}\, V_{tb}V_{tq^{'}}^{*}\, e_{q}\,  a_{9}(\mu)\, 
f_{V}\, m_{V}\, F_{1}^{B-P_1}(m_{V}^{2})\, (\varepsilon^{*}_{V}\cdot p_{B}),\\
P_{EW}^{VP_1}&=-\frac{3\sqrt{2}}{2}\, G_{F}\, V_{tb}V_{tq^{'}}^{*}\, e_{q}\, a_{9}(\mu)\, 
f_{P_1}\, m_{V}\, A_{0}^{B-V}(m_{P_1}^{2})\, (\varepsilon^{*}_{V}\cdot p_{B})\, .
\end{aligned}
\end{align}
In these equations, we use the superscripts $P_1 V, V P_1$ to distinguish cases 
in which the recoiling meson (the first particle of $P_1 V, V P_1$) is a pseudoscalar or a vector meson.
The quark in the Cabibbo-Kobayashi-Maskawa (CKM) matrix element is $q' =d,s$. $\varepsilon^{*}_{V}$ is the polarization vector 
of vector meson $V$. $r_{\chi}=\frac{m_{P_1}^2}{m_{q'}m_B}$ is the chiral factor of pseudo-scalar meson in the ``chiral enhanced" term 
of $P^{VP_1}$ amplitude. $f_{P_1}$, $f_{V}$ are the decay constants of the corresponding 
meson $P_1$ and $V$. $F_{1}^{B-P_1}$ and $A_{0}^{B-V}$ represent the vector form factors 
of $B_{(s)} \to P_1$ and $B_{(s)} \to V$ transitions. The $Q^2$ dependent of form factor is 
expressed in the dipole model as
\begin{equation}\label{eq:ffdipole}
F_{i}(Q^{2})
={F_{i}(0)\over 1-\alpha_{1}{Q^{2}\over m_{\rm pole}^{2}}+\alpha_{2}{Q^{4}\over m_{\rm pole}^{4}}},
\end{equation}
where $F_{i}$ represents form factor $F_{1}$ or $A_{0}$, and $m_{\rm pole}$ is the mass of the 
corresponding pole state, e.g., $B$ for $A_{0}$, and $B^{*}$ for $F_{1}$. 
The effective Wilson coefficients $ a_i ({\mu})$ can be calculated perturbatively. $\chi^{C^{(\prime)} 
(E,\, P,\, P_C^{(\prime)},\, P_A)}$ and $\phi^{C^{(\prime) }(E,\, P,\, P_C^{(\prime)},\, P_A)}$
denote the magnitude and associate phase of $C$ ($E$, $P$, $P_C$, and $P_A$ ) 
diagram, which are universal to be fitted globally from the experimental data.

We adopt the relativistic Breit-Wigner  line shape to describe 
$\rho$,\, $K^*$,\, $\omega$, and $\phi$ resonances, which is widely used in the experimental data 
analyses \cite{BaBar:2011vfx, BaBar:2012bdw, LHCb:2019sus}. 
The explicit expression of relativistic Breit-Wigner  line shape 
is in the following form:
\begin{align}\label{RBW}
L^{\mathrm{RBW}}(s)=\frac{1}{s-m_{V}^{2}+i m_{V} \Gamma_{V}(s)}\, ,
\end{align}
where the two-body invariant mass square is $s=m^2_{23}=(p_2+p_3)^2$ with
$p_2$ and $p_3$ denoting the 4-momenta of the collinearly moving mesons $P_2$ and $P_3$, 
respectively. The energy-dependent width of vector resonance $\Gamma_{V}(s)$ is defined by
\begin{align}\label{width}
\Gamma_{V}(s)=\Gamma_{0}\left(\frac{q}{q_{0}}\right)^{3}\left(\frac{m_{V}}{\sqrt{s}}\right) 
X^{2}\left(q\, r_{\mathrm{BW}}\right)\, .
\end{align}
The Blatt-Weisskopf barrier factor $X\left(q\, r_{\mathrm{BW}}\right)$ is given as~\cite{Blatt:1952ije}
\begin{equation}
X\left(q\, r_{\mathrm{BW}}\right)
=\sqrt{[1+\left(q_{0}\, r_{\mathrm{BW}}\right)^{2}]/[1+\left(q\, r_{\mathrm{BW}}\right)^{2}}]\, ,
\end{equation}
where $q=\frac{1}{2} \sqrt{\left[s-\left(m_{P_2}+m_{P_3}\right)^{2}\right]
 \left[s-\left(m_{P_2}-m_{P_3}\right)^{2}\right] / s}$ is the magnitude of the momentum of the final state $P_2$ or $P_3$ in the rest frame of 
resonance $V$, and $q_0$ is the value of $q$ when $s = m^2_{V}$.
When a pole mass locates outside the kinematics region, i.e., $m_{V}<m_{P_2}+m_{P_3}$,
$m_V$ will be replaced with an effective mass $m^{\mathrm{eff}}_{V}$, given by the ad hoc
formula \cite{Aaij:2014baa, Aaij:2016fma},
\begin{align}\label{massDstar}
 m_{V}^{\text {eff }}\left(m_{V}\right)=m^{\min }+\left(m^{\max }-m^{\min }\right)
 \left[1+\tanh \left(\frac{m_{V}-\frac{m^{\min }+m^{\max }}{2}}{m^{\max }-m^{\min }}\right)\right]\, ,
 \end{align}
 where $m^{\max }$ and $m^{\min }$ are the upper and lower boundaries of the kinematics region, respectively.
 The barrier radius $r_{\mathrm{BW}}=4.0 (\mathrm{GeV})^{-1}$ is for all resonances \cite{LHCb:2019sus}.
 The full widths of the resonances $\Gamma_{0}$,  together with their masses $m_V$,
 are taken from Particle Data Group~\cite{Workman:2022ynf} and listed in Table \ref{tab:mass and width}.
\begin{table}[tbhp]
\caption{Masses $m_V$ and full widths $\Gamma_0$   
of vector resonant states. }
\vspace{3mm}
\label{tab:mass and width}
\centering
\begin{tabular}{cccc}
\hline
Resonance ~~&~~ Line shape Parameters~~ &~~Resonance ~~&~~ Line shape Parameters
\\ \hline
$\rho(770)~~$  & $m_V\, =\, 775.26\,  \, \mathrm{MeV}$ &
$\omega(782)~~$  & $m_V\, =\, 782.65\,  \, \mathrm{MeV}$ \\
& $\Gamma_0\, =\, 149.1\, \, \mathrm{MeV}$
& &$\Gamma_0\, =\, 8.49\, \, \mathrm{MeV}$\\
$K^*(892)^+~~$  & $m_V\, =\, 891.66\,  \, \mathrm{MeV}$ 
&$K^*(892)^0~~$  & $m_V\, =\, 895.55\,  \, \mathrm{MeV}$\\
& $\Gamma_0\, =\, 50.8\, \, \mathrm{MeV}$
 & &$\Gamma_0\, =\, 47.3\, \, \mathrm{MeV}$\\
$\phi(1020)~~$ & $m_V\, =\, 1019.46\,  \, \mathrm{MeV}$ \\
&  $\Gamma_0\, =\, 4.25 \, \, \mathrm{MeV}$\\
\hline
\end{tabular}
\end{table}

The matrix element of strong decay $\left\langle  P_2 \left(p_{2}\right) P_3 \left(p_{3}\right) | V (p_V)\right\rangle$ 
is parametrized as a strong coupling constant $g_{ V P_2 P_3}$, which can be extracted from measured 
the partial decay widths $\Gamma_{V  \to P_2 P_3}$ through the relations
\begin{align}
\Gamma_{V \rightarrow P_{2} P_{3}}
=\frac{2}{3} \frac{p_{c}^{3}}{4 \pi m_{V}^{2}} g_{V \rightarrow P_{2} P_{3}}^{2}\, ,
\end{align}
where $p_c$ is the magnitude of pseudoscalar meson momentum in the rest frame of vector meson. The numerical results of $g_{\rho \rightarrow \pi^{+} \pi^{-}}$,
 $g_{K^{*} \rightarrow K^{+} \pi^{-}}$, and $g_{\phi \rightarrow K^{+} K^{-}}$ have already been 
 determined  from experimental data~\cite{Cheng:2013dua},
\begin{align}\label{gVPP}
&g_{\rho \rightarrow \pi^{+} \pi^{-}}=6.0\, ,
\quad g_{K^{*} \rightarrow K^{+} \pi^{-}}=4.59\, ,
\quad g_{\phi \rightarrow K^{+} K^{-}}=-4.54\, .
\end{align}
The other strong coupling constants can be derived from the relationships with the results in Eq.(\ref{gVPP} ) 
by using the quark model result \cite{Bruch:2004py},
$$g_{\rho \rightarrow K^{+} K^{-}}: g_{\omega \rightarrow K^{+} K^{-}}: g_{\phi \rightarrow K^{+} K^{-}}=1: 1:- \sqrt{2},$$
$$g_{\rho^{0} \pi^{+} \pi^{-}} = g_{\rho^+ \pi^0 \pi^+}\, , \, g_{\rho^{0} \pi^{0} \pi^{0}}= g_{\omega \pi^+ \pi^-} =0  \, ,$$
$$g_{\rho^{0} K^{+} K^{-}}=-g_{\rho^{0} K^{0} \bar{K}^{0}}=g_{\omega K^{+} K^{-}}=g_{\omega K^{0} \bar{K}^{0}}, \, 
g_{\phi K^{+} K^{-}}=g_{\phi K^{0} \bar{K}^{0}}\, .$$

Combing them together for quasi-two-body decay $B_{(s)}  \to  P_1 V \to  P_1 P_2 P_3$, 
we express the decay amplitudes of tree topological diagrams shown in Fig.\ref{tree} as
\begin{align}\label{eq:T}
\begin{aligned}
T^{(P_2 P_3)P_1}&=\left\langle P_2 \left(p_{2}\right) P_3 \left(p_{3}\right) \left|(\bar{u} b)_{V-A}\right| B (p_B) \right\rangle
\left \langle P_1 \left(p_{1}\right)\left|(\bar{q} u)_{V-A}\right| 0\right\rangle\\
&=\frac{\left\langle  P_2 \left(p_{2}\right) P_3 \left(p_{3}\right) | V (p_V)\right\rangle}{s-m_{V}^{2}+i m_{V} \Gamma_{V}(s)}
\left\langle V (p_V) \left|(\bar{u} b)_{V-A}\right| B(p_B) \right\rangle\left \langle P_1 \left(p_{1}\right)\left|(\bar{q} u)_{V-A}\right| 0\right\rangle\\
&=p_1 \cdot\left(p_{2}-p_{3}\right)\, \sqrt 2 G_{F} V_{u b} V_{u q^\prime}^{*} \, a_1({\mu})\, f_{P_1}\, m_{V} A_0^{B V} (m_{P_1}^2)\, \frac{g_{ V P_2 P_3}} {s-m_{V}^{2}+i m_{V} \Gamma_{V}(s)} \, ,\\
T^{P_1(P_2 P_3)}&=\left\langle P_2 \left(p_{2}\right) P_3 \left(p_{3}\right)  |(\bar{q} u)_{V-A} |0 \right \rangle
\left\langle P_1 (p_1)|(\bar u b)_{V-A}| B(p_B) \right \rangle \\
&=\frac{\left\langle  P_2 \left(p_{2}\right) P_3 \left(p_{3}\right) | V (p_V)\right\rangle}{s-m_{V}^{2}+i m_{V} \Gamma_{V}(s)}
\left\langle V (p_V) \left|(\bar{q} u)_{V-A}\right| 0 \right\rangle\left \langle P_1 \left(p_{1}\right)\left|(\bar{u} b)_{V-A}\right| B(p_B)\right\rangle\\
&=p_1 \cdot\left(p_{2}-p_{3}\right)\, \sqrt 2 G_{F} V_{u b} V_{u q^\prime}^{*} \, a_1({\mu})\, f_{V} m_{V} F_1^{B P_1} (s)\,  \frac{g_{ V P_2 P_3}} {s-m_{V}^{2}+i m_{V} \Gamma_{V}(s)} \, ,
\end{aligned}
\end{align}

\begin{align}\label{eq:C}
\begin{aligned}
C^{(P_2 P_3) P_1}&=\left\langle P_2 \left(p_{2}\right) P_3 \left(p_{3}\right) \left|(\bar{u} b)_{V-A}\right| B (p_B) \right\rangle
\left \langle P_1 \left(p_{1}\right)\left|(\bar{q} u)_{V-A}\right| 0\right\rangle\\
&=\frac{\left\langle  P_2 \left(p_{2}\right) P_3 \left(p_{3}\right) | V (p_V)\right\rangle}{s-m_{V}^{2}+i m_{V} \Gamma_{V}(s)}
\left\langle V (p_V) \left|(\bar{u} b)_{V-A}\right| B(p_B) \right\rangle\left \langle P_1 \left(p_{1}\right)\left|(\bar{q} u)_{V-A}\right| 0\right\rangle\\
&=p_1 \cdot\left(p_{2}-p_{3}\right)\, \sqrt 2 G_{F} V_{u b} V_{u q^\prime}^{*}  \chi^{C} \mathrm{e}^{i \phi^{C}}\, 
f_{P_1}\, m_{V} A_0^{B V} (m_{P_1}^2)\, \frac{g_{ V P_2 P_3}} {s-m_{V}^{2}+i m_{V} \Gamma_{V}(s)} \, ,\\
C^{P_1 (P_2 P_3) }&=\left\langle P_2 \left(p_{2}\right) P_3 \left(p_{3}\right) |(\bar{q} u)_{V-A} |0 \right \rangle
\left\langle P_1 (p_1)|(\bar u b)_{V-A}| B(p_B) \right \rangle \\
&=\frac{\left\langle  P_2 \left(p_{2}\right) P_3 \left(p_{3}\right)| V (p_V)\right\rangle}{s-m_{V}^{2}+i m_{V} \Gamma_{V}(s)}
\left\langle V (p_V) \left|(\bar{q} u)_{V-A}\right| 0 \right\rangle\left \langle P_1 \left(p_{1}\right)\left|(\bar{u} b)_{V-A}\right| B(p_B)\right\rangle\\
&=p_1 \cdot\left(p_{2}-p_{3}\right)\, \sqrt 2 G_{F} V_{u b} V_{u q^\prime}^{*}\, \chi^{C^{\prime}} \mathrm{e}^{i \phi^{C^{\prime}}}\, 
  f_{V} m_{V} F_1^{B P_1} (s)\,\frac{g_{ V  P_2 P_3}} {s-m_{V}^{2}+i m_{V} \Gamma_{V}(s)} \, ,\\
\end{aligned}
\end{align}

\begin{align}\label{eq:E}
\begin{aligned}
E^{P_1 (P_2 P_3)}&=\left\langle P_1 \left(p_{1}\right) P_2 \left(p_{2}\right) P_3 (p_3) \left|\mathcal{H}_{eff}\right| B(p_B) \right\rangle\\
&=\frac{\left\langle  P_2 \left(p_{2}\right) P_3 \left(p_{3}\right) | V (p_V)\right\rangle}{s-m_{V}^{2}+i m_{V} \Gamma_{V}(s)}
\left\langle V (p_V) P_1(p_1) \left|\mathcal{H}_{eff} \right| B(p_B)\right\rangle\\
&=p_1 \cdot\left(p_{2}-p_{3}\right)\, \sqrt 2 G_{F} V_{u b} V_{u q^\prime}^{*}\, \chi^{E} \mathrm{e}^{i \phi^{E}} \, 
f_{B} m_{V} \frac{f_{V} f_{P_1}}{f_{\pi}^2}\, \frac{g_{V  P_2 P_3}} {s-m_{V}^{2}+i m_{V} \Gamma_{V}(s)} \, ,
\end{aligned}
\end{align}
where $q^\prime=d, s$ and $p_V=p_2+p_3=\sqrt{s}$.
Similarly, the amplitudes of penguin diagrams shown in Fig.\ref{penguin}  can be expressed as 
\begin{align}\label{eq:P}
\begin{aligned}
P^{(P_2 P_3)P_1} &=-p_3 \cdot\left(p_{1}-p_{2}\right)\, 
 \sqrt 2 G_{F} V_{tb}V_{t q^\prime}^{*}[a_{4}(\mu)-\chi^{P} \mathrm{e}^{i\phi^{P}}r_{\chi}]\, 
 f_{P_3}\, m_{V} A_0^{B V} (m_{P_3}^2)\, \frac{g_{ V P_1 P_2}} {s-m_{V}^{2}+i m_{V} \Gamma_{V}(s)} \, ,
\\
P^{P_1(P_2 P_3)}&=-p_3 \cdot\left(p_{1}-p_{2}\right)\, 
 \sqrt 2 G_{F} V_{tb}V_{t q^\prime}^{*}\, a_{4}(\mu) \, 
 f_{V}\, m_{V} F_{1}^{B P_1} (s)\, \frac{g_{ V P_1 P_2}} {s-m_{V}^{2}+i m_{V} \Gamma_{V}(s)} \, ,
\\
P_A^{P_1(P_2 P_3)}&=-p_1 \cdot\left(p_{2}-p_{3}\right)\, \sqrt 2 G_{F} V_{tb}V_{t q^\prime}^{*}\, \chi^{P_A} \mathrm{e}^{i \phi^{P_A}}\, 
 f_{B}\, m_{V} \, \frac{f_{V} f_{P_1}}{f_{\pi}^2}\,  \frac{g_{ V P_2 P_3}} {s-m_{V}^{2}+i m_{V} \Gamma_{V}(s)} \, ,
\\
P_C^{(P_2 P_3) P_1} &=-p_1 \cdot\left(p_{2}-p_{3}\right)\, 
 \sqrt 2 G_{F} V_{tb}V_{t q^\prime}^{*}\, \chi^{P_C} \mathrm{e}^{i \phi^{P_C}}\, 
 f_{P_1}\, m_{V} A_0^{B V} (m_{P_1}^2)\, \frac{g_{ V P_2 P_3}} {s-m_{V}^{2}+i m_{V} \Gamma_{V}(s)} \, ,
\\
P_C^{P_1(P_2 P_3)}&=-p_1 \cdot\left(p_{2}-p_{3}\right)\,
 \sqrt 2 G_{F} V_{tb}V_{t q^\prime}^{*}\,  \chi^{P_C^{\prime}} \mathrm{e}^{i \phi^{P_C^{\prime}}}\, 
 f_{V}\, m_{V} F_{1}^{B P_1} (s)\, \frac{g_{ V P_2 P_3}} {s-m_{V}^{2}+i m_{V} \Gamma_{V}(s)} \, ,
 \\
P_{\mathrm{EW}}^{(P_2 P_3) P_1} &=-p_1 \cdot\left(p_{2}-p_{3}\right)\, 
 \frac{3\sqrt 2 }{2} G_{F} V_{tb}V_{t q^\prime}^{*}\, e_q   a_9(\mu)\, 
 f_{P_1}\, m_{V} A_0^{B V} (m_{P_1}^2)\, \frac{g_{ V P_2 P_3}} {s-m_{V}^{2}+i m_{V} \Gamma_{V}(s)} \, ,
 \\
P_{\mathrm{EW}}^{P_1(P_2 P_3)}&=-p_1 \cdot\left(p_{2}-p_{3}\right)\, 
 \frac{3\sqrt 2 }{2}G_{F} V_{tb}V_{t q^\prime}^{*}\,  e_q  a_9(\mu)\, 
 f_{V}\, m_{V} F_{1}^{B P_1} (s)\, \frac{g_{ V P_2 P_3}} {s-m_{V}^{2}+i m_{V} \Gamma_{V}(s)} \, ,
\end{aligned}
\end{align}
where the form factor $F_{1}^{B P_1}$ is dependent on invariant mass $s$ 
not a fixed value as in two-body decays. 

The matrix element of $B_{(s)}  \to  P_1 P_2 P_3$ can be written in the following form:
\begin{align}
\left\langle P_1 (p_{1})  P_2 (p_2) P_3 (p_{3}) \left| \mathcal{H}_{eff} \right| B_{(s)}(p_B) \right\rangle
= p_1 \cdot\left(p_{2}-p_{3}\right)\, \mathcal{A}(s)\, ,
\end{align}
where $\mathcal{A}(s)$ represents the summation of amplitudes in Eqs.(\ref{eq:T})-(\ref{eq:E}) 
and Eq.(\ref{eq:P}) with the prefactor $p_1 \cdot\left(p_{2}-p_{3}\right)$ taken out.
The differential width of $B_{(s)}  \to  P_1 P_2 P_3$ is
\begin{align}\label{dwidth}
\begin{aligned}
d \, \Gamma=&d\, s  \,  \frac{1}{(2 \pi)^3}\,
\frac{(\left|\mathbf{p_1} \| \mathbf{p}_{2}\right|\,)^3}{6 m_B^3}\, | \mathcal{A}(s)|^2\, , \\
\end{aligned}
\end{align}
where $|\mathbf{ p_{1}} |$ and $|\mathbf{ p_{2}} |$ denote the magnitudes of the
momentum $p_1$ and $p_2$, respectively. 
In the rest frame of the vector resonance, their expressions are
\begin{align}
\begin{aligned}
|\mathbf{ p_{1}} |=&\frac{1}{2\, \sqrt s}\,
\sqrt{ \left[(m_B^2-m_{P_1}^2)^2 \right] \, -2(m_B^2+m_{P_1}^2)^2 \,s +s^2}\, ,\\
|\mathbf{ p_{2}} |=&\frac{1}{2\, \sqrt s}\,
\sqrt{ \left[s-(m_{P_2}+m_{P_3})^2 \right] \, \left[s-(m_{P_2}-m_{P_3})^2 \right]}\, ,
\end{aligned}
\end{align}
where $|\mathbf{ p_{2}} |=q$ in Eq.(\ref{width}).

\section{Numerical Results of Branching Ratios}\label{sec:3}

In the numerical calculations, we need input parameters, such as  $m_V$, $\Gamma_0$, and $g_{V P_1 P_2}$ involved in strong interaction decay of vector mesons, which are listed 
in the previous section. For the CKM matrix elements needed 
in the weak transition calculation, 
we use Wolfenstein parametrization with parameters from global fit~\cite{Workman:2022ynf} as follows:
$$ \lambda=0.22650 \pm 0.00048,~~~A=0.790^{+0.017}_{-0.012},
~~~\bar \rho=0.141^{+0.016}_{-0.017},~~~\bar \eta=0.357 \pm 0.01.$$
The decay constants of light pseudoscalar mesons and vector mesons, and transition form factors 
of $B$-meson decays at recoil momentum square $Q^2=0$ are listed in Tables \ref{tab:decay constants} 
and \ref{tab:ff}, respectively. The decay constants of pseudoscalar mesons $\pi, K$, and $B$ are from Particle Data Group~\cite{Workman:2022ynf}. The decay constants of $B_s$ and the vector mesons, and 
all form factors are not measured by experiments and only calculated theoretically, 
such as in constitute quark model and light cone quark model 
\cite{Melikhov:2000yu,Geng:2001de,Lu:2007sg,Albertus:2014bfa}, 
covariant light front approach \cite{Cheng:2003sm, Cheng:2009ms, Chen:2009qk}, 
light-cone sum rules 
\cite{Bharucha:2015bzk,Ball:1998kk,Ball:1998tj,Ball:2001fp,Ball:2004ye,Ball:2004rg,Ball:2007hb,Charles:1998dr,Bharucha:2010im,Bharucha:2012wy,Khodjamirian:2006st,Khodjamirian:2011ub,Wang:2015vgv,Meissner:2013hya,Wu:2006rd,Wu:2009kq,Ivanov:2011aa,Ahmady:2014sva,Fu:2014uea,Gao:2019lta,Cui:2022zwm}, 
PQCD \cite{Li:2012nk,Wang:2012ab,Wang:2013ix,Fan:2013qz,Fan:2013kqa,Kurimoto:2001zj,Lu:2002ny,Wei:2002iu,Huang:2004hw}, and
lattice QCD \cite{Horgan:2013hoa,Dalgic:2006dt,Aoki:2013ldr}. 
We will use the same theoretical values 
as in the charmless two-body decays~\cite{Zhou:2016jkv}, 
with 5\% uncertainty kept for decay constants and 10\% for form factors.  
The dipole model parameters $\alpha_1, \alpha_2$ applied to describe the $Q^2$ dependence of form factors are  also listed in Table\ref{tab:ff}~\cite{Lu:2007sg,Wang:2012ab}.

\begin{table}[tbhp]
\caption{  Decay constants of light pseudoscalar mesons and vector mesons  (in units of MeV). }
\vspace{3mm}
\label{tab:decay constants}
\centering
\begin{tabular}{ccccccccccc|}
\hline
$f_{\pi}$ & $f_{K}$  &  $f_{B}$ & $f_{\rho}$ &$f_{K^*}$ & $f_{\omega}$&  $f_{\phi}$ &
\\ \hline
$130.2 \pm 1.7$ & $155.6 \pm 0.4$ & $190.9 \pm 4.1$ & $213 \pm 11$ & $220 \pm 11$ & $192\pm 10 $ & $225\pm11$ &
\\
\hline
\end{tabular}
\end{table}

\begin{table} [hbt]
\caption{Transition form factors at $Q^2=0$ and dipole model parameters used in this work. }\label{tab:ff}
\vspace{3mm}
\centering
\begin{tabular}{|c|c|c|c|c|c|c|c|c|c|c|c|c|c|}
\hline&
$~~F_{1}^{B\to\pi}~~$&
$~~F_{1}^{B\to K}~~$&
$~~F_{1}^{B\to \eta_q}~~$&
$~~A_{0}^{B\to \rho}~~$&
$~~A_{0}^{B\to \omega}~~$&
$~~A_{0}^{B\to K^*}~~$\\
\hline
$~F_i(0)~$&
0.28&
0.31&
0.21&
0.36&
0.32&
0.39\\
$\alpha_1$&
0.52&
0.54&
1.43&
1.56&
1.60&
1.51\\
$\alpha_2$&
0.45&
0.50&
0.41&
0.17&
0.22 &
0.14 \\
\hline
\end{tabular}
\end{table}

The effective Wilson coefficients $a_i(\mu)$ calculated at next-to-leading order can be found 
in~\cite{Li:2005kt}, $a_1(m_b/2)=1.054$, $a_4(m_b/2)=-0.04$, and $a_9(m_b/2)=-0.009$.
 The 14 nonpertubative  parameters of  topological diagrams 
$\chi^{C^{(\prime)}}, \chi^E, \chi^P,\chi^{P_C^{(\prime)}}, \chi^{P_A}$, and
associated phases $\phi^{C^{(\prime)}}, \phi^E, \phi^P,\phi^{P_C^{(\prime)}}, \phi^{P_A}$, extracted from experimental data by 
a global fit performed in Ref. \cite{Zhou:2016jkv}, together with   
 uncertainty are
\begin{align}\label{parameter}
\begin{aligned}
\chi^{C}=0.48 \pm 0.06,~~~&\phi^{C}=-1.58 \pm 0.08,\\
\chi^{C^{\prime}}=0.42 \pm 0.16,~~~&\phi^{C^{\prime}}=1.59\pm 0.17, \\
\chi^{E}=0.057\pm0.005,~~~&\phi^{E}=2.71\pm 0.13, \\
\chi^{P}=0.10\pm0.02,~~~&\phi^{P}=-0.61\pm 0.02, \\
\chi^{P_C}=0.048 \pm 0.003,~~~&\phi^{P_C}=1.56 \pm 0.08, \\
\chi^{P_C^{\prime}}=0.039\pm 0.003,~~~&\phi^{P_C^{\prime}}=0.68 \pm 0.08, \\
\chi^{P_A}=0.0059\pm0.0008,~~~&\phi^{P_A}=1.51\pm 0.09.
\end{aligned}
\end{align}


With all the inputs, the branching fractions of $B_{(s)}  \to  P_1 V \to  P_1 P_2 P_3$ 
can be obtained by integrating the differential width in Eq.(\ref{dwidth}) over the whole 
kinematics region. Our numerical results for the branching ratios of $B_{(s)}$ decays 
are collected in Tables \ref{Prho}-\ref{Pomega}, for decays 
$B_{(s)} \to P_1 \rho \to P_1 \pi \pi$, $B_{(s)} \to P_1 K^{*} \to P_1 K \pi$, 
$B_{(s)} \to P_1 \phi \to P_1 K\bar K$, and $B_{(s)} \to P_1 \rho, P_1 \omega \to P_1 K\bar K$, respectively. 
In these tables, we also list the intermediate resonances decays 
as well as the topological contributions represented by the corresponding symbols 
$T, C^{(\prime)}$ and so on. In our results, the first uncertainty is from the nonpertubative 
parameters in Eq.(\ref{parameter}), and the other two uncertainties are estimated with 
$10\%$ variations of form factors, $5\%$ variations of decay constants. One can see that 
the dominant source of uncertainties are from form factors. 

\subsection{Branching ratios of $B_{(s)} \to P_1 \rho \to P_1 \pi \pi$}

\begin{table}[!hb]
\caption{Branching ratios ($\times 10^{-6}$) of quasi-two-body decays 
$B_{(s)} \to P_1 \rho \to P_1 \pi \pi$ together with experimental data~\cite{Workman:2022ynf}.
The characters $T$, $C^{(\prime)}$, $E$, $P$, $P_C^{(\prime)}$, $P_A$ and $P_{\mathrm{EW}}$ 
representing the corresponding topological diagram contributions are also listed in the second column.}
 \label{Prho}
\begin{center}
\begin{tabular}{cccc}
 \hline \hline
{Modes}     &  Amplitudes & FAT results       &  Experiment   \\
\hline
  $B^- \to \pi^- (\rho^0\to)\pi^+ \pi^-$  & $T,C^{\prime}, P, P_A, P_\mathrm{EW}$ &$8.08 \pm 1.74 \pm 1.29 \pm 0.19 $   &$~~8.30\pm{1.20}~~$\\
  
     $B^- \to \pi^0(\rho^-\to)\pi^- \pi^0$ & $T, C, P, P_A, P_\mathrm{EW}$  &$12.70 \pm 0.71 \pm 2.24 \pm 1.12$  &$10.90\pm{1.40}$\\

     $\bar {B}^0 \to \pi^-(\rho^+\to)\pi^+ \pi^0$     &$T, E, P, P_A $         &$5.72 \pm 0.63 \pm 1.61 \pm 0.33$  &$8.40\pm1.10$ \\  
                             
  $\bar{B}^0 \to \pi^+(\rho^-\to)\pi^- \pi^0$    &$T, E, P $   &$12.10 \pm 0.91 \pm 3.14 \pm 1.27$  &$14.60\pm1.60$\\
      
     $\bar {B}^0 \to \pi^0(\rho^0\to)\pi^+ \pi^-$           &$C^{(\prime)}, E, P, P_A,  P_\mathrm{EW} $    &$1.23\pm 0.51\pm 0.08\pm0.16$  &$2.00\pm{0.50}$\\
                                   
                                 
                                                                                                  
 \hline 
   $B^- \to K^-(\rho^0\to)\pi^+ \pi^-$             &$T, C^{\prime}, P, P_\mathrm{EW} $~~~         &$3.41\pm0.25\pm0.80\pm0.04$  &$~~3.70\pm{0.50}~~$\\
  
    $B^- \to \bar{K}^0(\rho^-\to)\pi^- \pi^0$     &$P$     &$6.85\pm 0.47\pm1.55\pm0.07$  &$7.30^{+1.00}_{-1.20}$\\ 
                           
  $\bar{B}^0 \to K^-(\rho^+\to)\pi^+ \pi^0$            &$T, P$         &$7.42 \pm0.44\pm1.65\pm0.07$  &$7.00\pm{0.90}$\\
     
      $\bar {B}^0 \to \bar{K}^0(\rho^0\to)\pi^+ \pi^-$            &$C^{\prime}, P, P_\mathrm{EW}$         &$4.13\pm0.34\pm0.79\pm0.04$  &$3.40\pm{1.10}$\\
                            
  $\bar{B}_s^0 \to K^+(\rho^-\to)\pi^- \pi^0$           &$ T, P, P_A$         &$17.00\pm 0\pm3.50\pm 0.20$  &...\\
                                                                
  $\bar{B}_s^0 \to K^0(\rho^0\to)\pi^+ \pi^-$       &$C^{\prime}, P, P_C^{\prime}, P_A, P_\mathrm{EW}$         &$1.55\pm 1.10\pm0.31\pm 0.02$  &...\\                                
 \hline
   $B^- \to \eta(\rho^-\to)\pi^- \pi^0$            &$T, C, P, P_C, P_A, P_\mathrm{EW}$~~~        &$7.93\pm 0.48\pm1.43\pm0.07$  &$~~7.00\pm{2.90}~~$\\
     
      $B^- \to \eta^{\prime}(\rho^-\to)\pi^- \pi^0$   &$T, C, P, P_C, P_A, P_\mathrm{EW}$         &$5.81\pm 0.48\pm1.43\pm0.07$  &$9.70\pm{2.20}$\\
                                 
  $\bar{B}^0 \to \eta(\rho^0\to)\pi^+ \pi^-$            &$C^{(\prime)}, E, P, P_C^{(\prime)}, P_A,  P_\mathrm{EW}$         &$4.20 \pm 1.15\pm0.39\pm0.17$  &$<1.5$\\
   
    $\bar{B}^0 \to \eta^{\prime}(\rho^0\to)\pi^+ \pi^-$   &$C^{(\prime)}, E, P, P_C^{(\prime)}, P_A,  P_\mathrm{EW}$         &$3.09 \pm 0.77\pm0.29\pm0.12$ &$<1.3$\\
                                  
  $\bar{B}_s^0 \to \eta(\rho^0\to)\pi^+ \pi^-$          &$C^{\prime}, E, P_C^{\prime}, P_\mathrm{EW}$         &$0.11 \pm0.02\pm0.02\pm0.003$  &...\\
                                                                                      
  $\bar{B}_s^0 \to \eta^{\prime}(\rho^0\to)\pi^+ \pi^-$ &$C^{\prime}, E, P_C^{\prime}, P_\mathrm{EW}$         &$0.34 \pm 0.07\pm0.05\pm0.01$  &...\\                                
 \hline \hline
\end{tabular}
\end{center}
\end{table}

Comparing the results shown 
in Table \ref{Prho} with the corresponding ones calculated  in the previous work ~\cite{Zhou:2016jkv}, we find that the branching ratios of quasi-two-body decay $B_{(s)} \to P_1 \rho \to P_1 \pi \pi$ are a little smaller than that of the direct   two-body decays 
$B_{(s)} \to P_1 \rho$. It can be attributed to the not narrow 
resonance width $\Gamma_\rho$. We cannot apply the narrow width approximation to 
factorize quasi-two-body decay formula together with $\mathcal{B}(\rho \to \pi \pi ) \sim 100\%$
to obtain the precise results of two-body decays. 

 From Table \ref{Prho}, it is easy to see that our results are well consistent with 
the experimental data shown in the last column of the Table.
For the decay modes $B \to \pi (\rho \to) \pi \pi $, 
$B \to \eta^{(\prime)} (\rho \to) \pi \pi $ with color favored $T$ diagram dominated contribution, and the penguin diagram $P$ dominated modes
$B_{(s)} \to K (\rho \to) \pi \pi $, the perturbation calculation is reliable, since these kinds of diagrams are proved to be factorizable to all orders of $\alpha_s$. Our results  are in good agreement with those calculated 
under the QCD factorization \cite{Cheng:2013dua} and that in PQCD approach~\cite{Li:2016tpn} within errors.

For the decay modes with dominated color suppressed diagrams $C^{(\prime)}$ contributions, such as  $ \bar {B}^0 \to \pi^0 (\rho^0 \to) \pi \pi $ and $\bar{B}^0 \to \eta^{(\prime)} (\rho^0 \to ) \pi \pi $, the theoretical results are sensitive to the power corrections and next-to-leading order contributions. 
With the magnitude and phase of this $C^{(\prime)}$ diagram given in Eq.(\ref{parameter}), our results of branching ratios are consistent with experimental data  
and results calculated 
under the  QCD factorization approach \cite{Cheng:2013dua}. On the contrary, the calculations of these decay modes  are only done to the leading order in the perturbative QCD approach with one order magnitude smaller 
branching ratio~\cite{Li:2016tpn}.


\subsection{Branching ratios of $B_{(s)} \to P_1 K^{*} \to P_1 K \pi$}

\begin{table}[tbhp]
\caption{Branching ratios ($\times 10^{-6}$) in FAT approach of  quasi-two-body decays 
$B_{(s)} \to P_1 K^{*} \to P_1 K \pi$, together with experimental data or QCD factorization results.
The characters $T$, $C^{(\prime)}$, $E$, $P$, $P_C^{(\prime)}$, $P_A$ and $P_{\mathrm{EW}}$ 
representing the corresponding topological diagram contributions are also listed in the second column.}
 \label{PKstar}
\begin{center}
\begin{tabular}{cccc}
 \hline \hline
Decay Modes     &  Amplitudes & FAT  results       &  Experiment/QCDF   \\
\hline
$B^- \to \pi^-(\bar{K}^{*0}\to)K^- \pi^+$     &~~~$P, P_A$~~~     &$6.91\pm 0.66 \pm1.19\pm0.67$    &$~~7.2\pm 0.4\pm 0.7^{+0.3}_{-0.5}$ \footnote{Experimental data from BABAR.}\\

  &  &  &$6.45 \pm 0.43 \pm 0.48_{-0.35}^{+0.25}$    \footnote{Experimental data from Belle.}\\

 $B^- \to \pi^0(K^{*-}\to)K^- \pi^0$      &~~~$T, C, P, P_A, P_\mathrm{EW}$~~~     &$1.91 \pm 0.16 \pm 0.29 \pm0.16$    &$~~2.7\pm0.5\pm0.4~^a$\\
 
 $\bar{B}^0 \to \pi^+(K^{*-}\to)K^- \pi^0$         &~~~$T, P,P_A$~~~     &$2.57\pm0.25\pm0.44\pm0.25 $    &$2.7 \pm 0.4 \pm 0.3~^a$\\
  
     &   & &$4.9_{-1.5-0.3-0.3}^{+1.5+0.5+0.8}~^b$\\
  
  $\bar{B}^0 \to \pi^0(\bar{K}^{*0}\to)K^- \pi^+$       &~~~$C, P, P_A, P_\mathrm{EW}$~~~     &$2.33\pm0.27\pm0.44\pm0.26$    &$~~2.2\pm 0.3 ~\pm 0.3~^a$\\
  
   &   &  &$< 2.3~^b$\\                              
                                                        
  $\bar{B}_s^0 \to \pi^0(K^{*0}\to)K^- \pi^+$        &~~~$C, P, P_\mathrm{EW}$~~~     &$0.90\pm0.18\pm0.17\pm0.003$   &...\\
\hline
 $B^- \to K^-(K^{*0} \to)K^+\pi^-$         &~~~$P, P_A$~~~     &$0.40\pm0.04\pm0.07\pm0.04$  &$0.22_{-0.00-0.04-0.01}^{+0.00+0.04+0.01}$ \footnote{Results from QCD factorization approach
  \cite{Cheng:2013dua, Cheng:2014uga} .}\\
  
   $B^- \to K^0(K^{*-}\to)K^- \pi^0$  &~~~$P$~~~     &$0.14\pm0.01\pm0.03\pm0.001$    &...\\
  
   $\bar{B}^0 \to K^0(\bar{K}^{*0}\to)K^- \pi^+$         &~~~$P$~~~     &$0.27\pm0.02\pm0.05\pm0.002$    &...\\

    $\bar{B}^0 \to \bar{K}^0(K^{*0}\to)K^- \pi^+$        &~~~$P, P_A$~~~     &$0.37\pm0.04\pm0.06\pm0.04$    &$0.20_{-0.00-0.03-0.00}^{+0.00+0.04+0.00}~^c$\\
 
   $\bar{B}_s^0 \to K^+(K^{*-}\to)K^- \pi^0$       &~~~$T, E, P, P_A$~~~     &$2.76\pm0.44\pm0.31\pm0.29$    &...\\                                           

  $\bar{B}_s^0 \to K^0(\bar{K}^{*0}\to)K^- \pi^+$       &~~~$P, P_A$~~~     &$6.36\pm0.98\pm0.82\pm0.66$   &$3.8_{-0.0-0.7-0.0}^{+0.0+0.8+0.0} ~^c$ \\   
                                  
  $\bar{B}_s^0 \to \bar{K}^0(K^{*0}\to)K^+ \pi^-$         &~~~$P$~~~     &$4.28\pm0.26\pm0.84\pm0.04$    &$1.5_{-0.0-0.9-0.0}^{+0.0+2.4+0.0}~^c$\\
  \hline
     $B^- \to \eta(K^{*-}\to)K^- \pi^0$            &$T, C, P, P_C, P_A, P_\mathrm{EW}$~~~        &$5.35\pm0.37\pm0.74\pm0.19$  &
     ...\\
     
      $B^- \to \eta^{\prime}(K^{*-}\to)K^- \pi^0$            &$T, C, P, P_C, P_A, P_\mathrm{EW}$~~~        &$1.02\pm0.35\pm0.12\pm0.06$  &...\\
      
      $\bar{B}^0\to \eta(\bar{K}^{*0}\to)K^-\pi^+$            &$C, P, P_C, P_A, P_\mathrm{EW}$~~~        &$11.3
\pm0.75\pm1.53\pm0.41$  &...\\
      
      $\bar{B}^0 \to \eta^{\prime}(\bar{K}^{*0}\to)K^-\pi^+$            &$C, P, P_C^{(\prime)}, P_A, P_\mathrm{EW}$~~~        
      &$2.01\pm0.74\pm0.21\pm0.12$  &...\\
   
      $\bar{B}_s^0\to \eta(K^{*0}\to)K^+\pi^-$            &$C, P, P_C, P_A, P_\mathrm{EW}$~~~        &$0.69\pm0.13\pm0.11\pm0.02$  &...\\ 
      
        $\bar{B}_s^0\to \eta^{\prime}(K^{*0}\to)K^+\pi^-$            &$C, P, P_C, P_A, P_\mathrm{EW}$~~~        &$1.17\pm0.14\pm0.16\pm0.04$  &...\\                                                                              
 \hline \hline
\end{tabular}
\end{center}
\end{table}

In Table  \ref{PKstar}, we only list quasi-two-body decays 
$B_{(s)} \to P_1 K^{*} \to P_1 K \pi$ through strong decays $K^{*0} \to  K^+ \pi^-$, 
$ \bar{K}^{*0} \to  K^- \pi^+$, and $ K^{*-} \to  K^- \pi^0$  
with a pair of $u \bar u$ from the sea.  
The branching ratios for $B_{(s)} \to P_1 K^{*} \to P_1 K \pi$ via sea quark pair $d \bar d$ strong decay
$ K^{*0} \to  K^0 \pi^0$, $ \bar{K}^{*0} \to  \bar{K}^0 \pi^0$, and $ K^{*-} \to  \bar{K}^0 \pi^-$
 can be calculated under the narrow width approximation according to
 the isospin conservation relationship for strong interaction, 
 \begin{eqnarray}
 \mathcal{B}(\bar{K}^{*0} \to K^- \pi^+) &=& 2\, \mathcal{B} (\bar{K}^{*0} \to \bar{K}^0 \pi^0),\\
 \mathcal{B} (K^{*-} \to \bar{K}^0 \pi^-)&=& 2\, \mathcal{B}(K^{*-} \to K^- \pi^0).
 \end{eqnarray}
 The experimental data listed in the last column are measured by BABAR
 ~\cite{BaBar:2008lpx, BaBar:2011cmh, BaBar:2009jov, BaBar:2011vfx}
and Belle~\cite{Belle:2005rpz, Belle:2004khm} experiments. As we have considered the 
power corrections from ``chiral enhanced" term, penguin annihilation contribution
and EW-penguin diagram for the penguin-dominated modes $B_{u,d} \to \pi K^{*} \to \pi K \pi$, 
our results are well consistent with experimental data, and larger by a factor of $2 \sim 4$ than 
QCD factorization approach~\cite{Cheng:2013dua}. 
 
The decay channels $B_{u,d} \to K K^{*} \to K K \pi$ listed in  Table  \ref{PKstar} are CKM suppressed ($|V_{td}/V_{ts}|$) comparing with decay $B_{u,d} \to \pi K^{*} \to \pi K \pi$,
  so that the branching ratios of the former are one order magnitude smaller than the latter. Since there are not experimental measurements 
of these decays and $B_{s} \to K K^{*} \to K K \pi$, we show in the last column the results 
 of QCD factorization approach~\cite{Cheng:2013dua, Cheng:2014uga} for comparison.
Similarly with the case happened in $B_{u,d} \to \pi K^{*} \to \pi K \pi$ decays, the results of 
$B_{u,d} \to K K^{*} \to K K \pi$ as well as penguin-dominated decays $B_{s} \to K K^{*} \to K K \pi$ 
calculated in FAT approach are larger by a factor of $2 \sim 3$ than those in QCD factorization 
approach~\cite{Cheng:2013dua}.

We also give the numerical results of  branching ratios of   decay modes $B_{(s)} \to \eta K^{*} \to \eta K \pi$, shown in Table  \ref{PKstar}.
 These decays have not been measured by experiments from the three-body Dalitz plot analyses.  However the corresponding two-body decays have been measured experimentally.   Applying the narrow width approximation for the $K^*$ resonance, we have 
$\Gamma\left(B_{(s)} \to   \eta K^{*} \to \eta K \pi \right)=
\Gamma(B \rightarrow \eta K^{*})  \mathcal{B}\left(K^{*} \to K \pi\right)$. 
 Utilizing the isospin symmetry relationship  
 \begin{eqnarray}
 \mathcal{B}\left(K^{* 0} \rightarrow K^{+} \pi^{-}\right)
 =\mathcal{B}\left(K^{*-} \rightarrow \bar{K}^0 \pi^{-}\right)
 =2\, \mathcal{B}\left(K^{*-} \rightarrow K^{-} \pi^0\right),
 \end{eqnarray}
we can obtain the branching ratios for two-body decays $B \to   \eta K^{*} $,
\begin{eqnarray}
\mathcal{B} \left(B^- \to \eta K^{*-} \right) & =&(16.05\pm1.11\pm2.22\pm0.57)\times 10^{-6},\\
\mathcal{B} \left(B^- \to \eta^{\prime} K^{*-}\right)& =&(3.06\pm1.05\pm0.36\pm0.18)\times 10^{-6},\\
\mathcal{B} \left(\bar{B}^0 \to \eta \bar{K}^{*0} \right) &=&(16.95\pm 1.13\pm2.30\pm0.62) \times 10^{-6},\\  
\mathcal{B} \left(\bar{B}^0 \to \eta^{\prime} \bar{K}^{*0} \right) &=&(3.02\pm1.11\pm0.32\pm0.18)\times 10^{-6}.
\end{eqnarray}
These results agree well with the experimental measurements of these two-body decays and the
ones given in FAT approach~\cite{Zhou:2016jkv}.

\subsection{Branching ratios of $B_{(s)} \to P_1 \phi \to P_1 K\bar K$}

\begin{table}[tbhp]
\caption{Branching ratios  of  quasi-two-body decays  $B_{(s)} \to P_1 \phi \to P_1 K\bar K$ from FAT approach and that from PQCD approach.
The characters  $C$,   $P$, $P_C^{(\prime)}$, $P_A$ and $P_{\mathrm{EW}}$ 
representing the corresponding topological diagram contributions are also listed in the second column.}
 \label{Pphi} 
 \vspace{0.5cm}
 \newsavebox{\tablebox}
\begin{lrbox}{\tablebox}
\centering
\begin{tabular}{cccc}
 \hline \hline
{Modes}     &  Amplitudes  &  ${\mathcal B}_{\rm {FAT}} $   &  ${\mathcal B}_{\rm {PQCD}} $   \\
\hline
 $B^-\to  \pi^- (\phi \,\to) K^+K^- $    &$P_C^{\prime}, P_\mathrm{EW}$\;
      & $1.47\pm0.19\pm0.30\pm0.15 \times 10^{-7}$  & $3.58\pm1.17\pm1.87\pm0.34 \times 10^{-9}$ \\
      
 $\quad~ \to  \pi^- (\phi \,\to) K^0 \bar{K}^0 $    &
      & $1.03\pm0.13\pm0.21\pm0.10 \times 10^{-7}$ &$2.47\pm0.81\pm1.30\pm0.24\times 10^{-9}$   \\
    
$\bar{B}^0 \;\to \pi^0 (\phi\;\to) K^+K^- $\    &$P_C^{\prime}, P_\mathrm{EW}$\;
      & $6.82\pm0.87\pm1.37\pm0.68 \times 10^{-8}$ &$1.74\pm0.53 \pm0.91\pm0.14\times 10^{-9}$ \\  
      
 $\quad~\to \pi^0 (\phi\;\to) K^0 \bar{K}^0 $\    &\;
      & $4.78\pm0.61\pm0.96\pm0.48 \times 10^{-8}$ &$1.20\pm0.37\pm0.63\pm0.10\times 10^{-9}$   \\   

$\bar{B}_s^0 \;\to \pi^0 (\phi\;\to) K^+K^- $ &$C, P_\mathrm{EW}$\;
      & $1.44 \pm0.13\pm0.29\pm0.004\times 10^{-7}$ & $9.11\pm2.03\pm0.14\pm0.61\times 10^{-8}$    \\  
      
      $ \quad~\to \pi^0 (\phi\;\to) K^0 \bar{K}^0$ &
      & $1.00\pm0.09\pm0.20\pm0.003 \times 10^{-7}$ &$6.30\pm1.40\pm0.10\pm0.43\times 10^{-8}$   \\    
    \hline      
 $B^-\to  K^- (\phi\!\to )K^+K^- $    &$P, P_C^{\prime}, P_A, P_\mathrm{EW}$\;
      & $4.53\pm1.00\pm0.38\pm0.52 \times 10^{-6}$ & $4.03\pm0.67\pm0.49\pm0.15\times 10^{-6}$    \\
      
      $\quad~\to  K^- (\phi\!\to ) K^0 \bar{K}^0 $    &
      & $3.16\pm0.70\pm0.26\pm0.37\times 10^{-6}$ &$2.79\pm0.46\pm0.34\pm0.11\times 10^{-6}$  \\
 
   $\bar{B}^0 \;\to  \bar{K}^0 (\phi \to )K^+K^- $\    &$P, P_C^{\prime}, P_A, P_\mathrm{EW}$\;
      & $4.20\pm0.93\pm0.35\pm0.48 \times 10^{-6}$ &$3.62\pm0.64\pm0.59\pm0.19\times 10^{-6}$   \\  
      
      $\quad~ \to  \bar{K}^0 (\phi \to )K^0 \bar{K}^0 $\    &
      & $2.94 \pm0.65\pm0.24\pm0.34\times 10^{-6}$ &$2.50\pm0.44\pm0.41\pm0.13 \times 10^{-6}$   \\     
  
     $\bar{B}_s^0 \;\to   K^0 (\phi \to) K^+K^- $ &$P, P_C^\prime, P_\mathrm{EW}$\;
      & $1.84\pm0.24\pm0.31\pm0.035 \times 10^{-7}$ &$8.34\pm0.48\pm0.94\pm2.07\times 10^{-8}$   \\     
      
      $ \quad~\to   K^0 (\phi \to) K^0 \bar{K}^0 $ &
      & $1.28\pm0.16\pm0.22\pm0.02 \times 10^{-7}$ & $5.76\pm0.33\pm0.65\pm1.44 \times 10^{-8}$    \\   
     
   \hline     
    $\bar{B}^0 \;\to \eta (\phi\;\to) K^+K^- $\    &$P_C^{\prime}, P_\mathrm{EW}$\;
      & $4.10 \pm0.52\pm0.82\pm0.41\times 10^{-8}$  & ...  \\ 
      
    $\quad~\to \eta (\phi\;\to) K^0 \bar{K}^0 $\    &
      & $2.87\pm0.37\pm0.57\pm0.29 \times 10^{-8}$ & ...    \\  
  
    $\bar{B}^0 \;\to \eta^{\prime} (\phi\;\to) K^+K^- $\    &$P_C^{\prime}, P_\mathrm{EW}$\;
      & $2.75 \pm0.35\pm0.55\pm0.27\times 10^{-8}$  & ...   \\  
    
    $\quad~ \to \eta^{\prime} (\phi\;\to) K^0 \bar{K}^0 $\    &
      & $1.92\pm0.25\pm0.38\pm0.19 \times 10^{-8}$ & ...   \\  

     $\bar{B}_s^0 \;\to \eta (\phi\;\to) K^+K^- $ &$C, P, P_C, P_A, P_\mathrm{EW}$\;
      & $4.14\pm2.50\pm2.77\pm1.21 \times 10^{-7}$  & ...   \\    
      
      $\quad~ \to \eta (\phi\;\to) K^0 \bar{K}^0$ &
      & $2.88\pm1.75\pm1.92\pm0.84 \times 10^{-7}$  & ...   \\   
       
        $\bar{B}_s^0 \;\to \eta^{\prime} (\phi\;\to) K^+K^- $ &$C, P, P_C, P_A, P_\mathrm{EW}$\;
      & $6.94\pm1.65\pm0.52\pm0.58 \times 10^{-6}$  & ...   \\    
      
       $\quad~ \to \eta^{\prime} (\phi\;\to) K^0 \bar{K}^0 $ &
      &  $4.83 \pm1.14\pm0.36\pm0.40\times 10^{-6}$  & ...   \\    
    \hline \hline
\end{tabular}
\end{lrbox}
 \scalebox{0.9}{\usebox{\tablebox}}
\end{table}

We show the branching fraction results of $B_{(s)} \to P_1 ( \phi \to) K\bar K$ decays in Table \ref{Pphi} 
together with the results from PQCD approach~\cite{Fan:2020gvr} in the last column.
For penguin diagram $P$ dominated decay modes, $B_{u,d} \to K ( \phi \to) K\bar K$,
the results in FAT approach are consistent with the ones in PQCD approach. As discussed in the previous section, this kind of decays are dominated by reliable perturbation contributions, which agree well with the experimental 
data $\mathcal{B}(B^- \to K^- ( \phi \to) K^+ K^-) =(4.48 \pm 0.22^{+0.33}_{-0.24})\times 10^{-6}$ \cite{BaBar:2012iuj} and 
$\mathcal{B}(B^- \to K^- ( \phi \to) K^+ K^- )=(4.72 \pm 0.45 \pm 0.35_{-0.22}^{+0.39})\times 10^{-6}$  \cite{Belle:2004drb}. 
The decay modes $B_{s} \to K ( \phi \to) K\bar K$ are CKM suppressed ($|V_{td}/V_{ts}|$) comparing with $B_{u,d} \to K ( \phi \to) K\bar K$ decays, so that 
the branching ratios of the former are one order smaller than the latter. It is easy to see from Table \ref{Pphi} that branching ratios of all the CKM suppressed decay modes 
from the FAT approach are larger that that from   PQCD approach. 
It can be attribute to the missing  power corrections and next-to-leading order corrections to the color suppressed penguin diagram  $P_C^\prime$ in PQCD approach.

Summing over the  two quasi-two-body decay modes with same weak decay
but different subsequent strong decay, i.e. $B_{(s)} \to  P_1 (\phi \,\to) K^+K^- $ 
and $B_{(s)} \to  P_1 (\phi \,\to) K^0 \bar{K}^0$,
we should obtain results of the corresponding two-body decays $B_{(s)} \to  P_1 \phi \,$ 
by applying the narrow width approximation.
 For instance, the summation of the results of 
$\mathcal {B} (B^-\to  \pi^- (\phi \,\to) K^+K^-)$ and 
$\mathcal{B}(B^-\to   \pi^- (\phi \,\to) K^0 \bar{K}^0 )$ 
in Table~\ref{Pphi} is in agreement with the two-body decay value 
$\mathcal{B}(B^-\to  \pi^- \phi )=(2.80 \pm 0.04 \pm 0.55 \pm 0.03)\times 10^{-7}$ in~\cite{Zhou:2016jkv}
within the error bar.

  
\subsection{The virtual effects of $B_{(s)} \to \pi, K ( \rho, \omega \to)K\bar K$}\label{subBWT}

\begin{table}[tbhp]
\caption{Comparison of results from FAT and PQCD approach for the virtual effects of $B_{(s)} \to \pi, K ( \rho, \omega \to)K \bar K$ decays,
 happened when the pole masses of $\rho, \omega $ are smaller than the invariant mass of $K \bar K$.}
 \label{Pomega}
\begin{center}
\begin{tabular}{cccc}
 \hline \hline
{Modes}     &  ${\mathcal B}_{\rm {FAT}} $   &  ${\mathcal B}_{\rm {PQCD}} $       \\
\hline
 $B^-\to \pi^0 (\rho^-\to) K^- K^0$\;    
 
       &$1.02\pm0.05\pm0.19\pm0.09\times10^{-7} $  &$2.01^{+0.38+0.29+0.24+0.10+0.06}_{-0.35-0.26-0.20-0.07-0.06}\times10^{-8}$
         \\
 $B^-\to \pi^- (\rho^0\to) K^+K^-$\;    
 
       &$5.48\pm1.40\pm0.85\pm0.16\times10^{-8}$ &$1.43^{+0.26+0.19+0.11+0.06+0.04}_{-0.25-0.17-0.10-0.05-0.04}\times10^{-7}$ 
         \\ 
  $B^-\to \pi^- (\omega\;\to) K^+K^-$\;    
 
       &$4.48\pm1.15\pm0.70\pm0.13\times10^{-8}$ &$4.21^{+1.67+1.03+0.08+0.21+0.14}_{-1.34-0.96-0.08-0.17-0.14}\times10^{-8}$ 
          \\   

  $\bar{B}^0\to \pi^+ (\rho^- \to) K^- K^0$\;    

       &  $1.10\pm0.07\pm0.27\pm0.11 \times10^{-7}$&$1.02^{+0.21+0.28+0.14+0.06+0.03}_{-0.17-0.25-0.13-0.05-0.03}\times10^{-7}$ 
          \\   
 $\bar{B}^0\to \pi^- (\rho^+\to) K^+ \bar{K}^0$\;    

       &$3.51\pm0.39\pm0.99\pm0.20 \times10^{-8}$  &$9.59^{+3.25+1.96+0.22+0.46+0.29}_{-2.90-1.88-0.19-0.33-0.29}\times10^{-8}$ 
          \\                
 $\bar{B}^0\to \pi^0\; (\rho^0\,\to) K^+K^-$\;    

       &$7.54\pm3.39\pm0.44\pm1.03 \times10^{-9}$ &$1.47^{+0.96+0.53+0.19+0.13+0.04}_{-0.78-0.49-0.14-0.07-0.04}\times10^{-9}$
          \\     

 $\bar{B}^0\to \pi^0\; (\omega\;\,\to) K^+K^-$\;    

       &$1.60 \pm0.67\pm0.18\pm0.11 \times10^{-8}$ 
       &  $4.96^{+0.73+1.25+0.63+0.24+0.17}_{-0.87-1.36-0.65-0.22-0.17}\times10^{-9}$ \\    
       

                     
       
\hline       
 $B^-\to \bar{K}^0 (\rho^- \to) K^- K^0$\;    

       &$4.16\pm0.25\pm0.84\pm0.04\times10^{-8}$ 
       & $2.21^{+0.51+0.51+0.34+0.10+0.07}_{-0.45-0.46-0.29-0.08-0.07}\times10^{-7}$  \\    
 $B^-\to K^- (\rho^0 \to) K^+K^-$\;    

       &$2.13 \pm0.14\pm0.43\pm0.02 \times10^{-8}$
       &  $5.15^{+0.91+0.99+0.69+0.25+0.16}_{-0.85-0.98-0.66-0.21-0.16}\times10^{-8}$  \\      
 $B^-\to K^- (\omega\;\to) K^+K^-$\;    

       &$5.09 \pm0.84\pm0.95\pm0.47 \times10^{-8}$ \;
       &  $8.92^{+1.67+2.33+1.19+0.43+0.30}_{-1.47-2.18-1.07-0.34-0.30}\times10^{-8}$  \\                                                
 $\bar{B}^0\to K^- (\rho^+\!\to) K^+ \bar{K}^0$\;    

       &$4.51\pm0.24\pm0.91\pm0.04 \times10^{-8}$ 
       &  $1.77^{+0.30+0.41+0.27+0.08+0.05}_{-0.25-0.39-0.25-0.06-0.05}\times10^{-7}$ \\            
 $\bar{B}^0\to \bar{K}^0\; (\rho^0\to)K^+K^-$\;    

       &$2.71\pm0.24\pm0.45\pm0.06\times10^{-8}$ 
       &  $5.44^{+0.88+1.26+0.82+0.24+0.17}_{-0.81-1.19-0.76-0.18-0.17}\times10^{-8}$   \\      
 $\bar{B}^0\to \bar{K}^0\; (\omega\;\to) K^+K^-$\;    

       &$3.81\pm0.71\pm0.81\pm0.41\times10^{-8} $ 
       &  $5.99^{+1.15+1.60+0.88+0.22+0.20}_{-0.96-1.39-0.75-0.19-0.20}\times10^{-8}$ \\      
 $\bar{B}_s^0\to K^+ (\rho^-\to) K^-  K^0$\;    

       &$1.40 \pm0.003\pm0.28\pm0.14\times10^{-7}$ 
       &   $2.04^{+0.03+0.43+0.22+0.11+0.06}_{-0.02-0.41-0.21-0.09-0.06}\times10^{-7}$ \\            
 $\bar{B}_s^0\to K^0\; (\rho^0\to) K^+K^-$\;    

       &$1.29\pm0.90\pm0.25\pm0.13 \times10^{-8}$ 
       &  $1.03^{+0.63+0.19+0.18+0.08+0.03}_{-0.45-0.17-0.16-0.05-0.03}\times10^{-9}$  \\    
 $\bar{B}_s^0\to K^0\; (\omega \;\to) K^+K^-$\;    

       &$1.09 \pm0.72\pm0.20\pm0.11\times10^{-8}$ 
       & $1.39^{+0.68+0.17+0.12+0.07+0.05}_{-0.57-0.14-0.14-0.07-0.05}\times10^{-9}$   \\    
\hline\hline
\end{tabular}
\end{center}
\end{table}

The decay modes $B_{(s)} \to \pi, K (\rho, \omega \to ) K \bar K$
represent a category of decays, whose pole masses of the resonances $\rho, \omega$ 
smaller than the threshold mass of producing $ K \bar K$ pair. They can only happen with 
the virtual effect, which is also called the Breit-Wigner tail  effect.
In Table \ref{Pomega}, we collect these virtual effects in FAT approach,
together with the results in PQCD approach \cite{Wang:2020nel} for comparison. 
Here we only show subprocesses $ \rho^0, \omega \to K^+ K^-$ without 
$ \rho^0, \omega \to K^0 \bar K^0$ due to the tiny mass difference between $K^+ K^-$ and $K^0 \bar{K}^0$. 
The virtual effects of $B_{(s)} \to \pi, K ( \rho \to)K\bar K$ 
are approximately $1\sim2$ order smaller than the dominated contribution of 
$B_{(s)} \to \pi, K ( \rho \to) \pi \pi$ in Table~\ref{Prho}. Unlike the on-shell resonance contributions that mostly give the similar contributions between the FAT and PQCD approaches, the virtual contributions give quite different results between the two approaches. Since the color suppressed tree diagram $C^{(\prime)}$ is calculated in the 
PQCD approach only to the leading order, its size is significantly smaller that that fitted from experimental data 
in the FAT approach~\cite{Zhou:2016jkv}. As a result, 
 the off-shell effects in $C^{(\prime)}$ dominated decay modes, $\bar{B}^0\to \pi^0\; (\rho^0,\omega\,\to) K^+K^-$, 
 $\bar{B}_s^0\to K^0\; (\rho^0,\omega \;\to) K^+K^-$ using FAT approach are nearly one order magnitude 
 larger than that of PQCD approach. 
For another example, the hierarchy between the magnitude of virtual contributions  for 
quasi-two-body decays $B^-\to \pi^0 ( \rho^- \to)K^- K^0$ and $B^-\to \pi^- ( \rho^0 \to)K^+ K^-$
is opposite in the PQCD and the FAT approaches,
which need further experimental measurements to check.

Comparing the results between Tables \ref{Prho} and \ref{Pomega}, one can see that the off-shell effects in $B \to \pi, K (\rho \to) K\bar K$ are small, that  is only about 1\% of the on-resonance contribution. However
 from Table \ref{Pomega}, one can see that the off-shell effect from the ground state $\rho(770)^0$ in $B^-\to \pi^- ( \rho^0 \to)K^+ K^-$ decay  in FAT approach
is  at the same order as the branching ratio
\begin{eqnarray}
 {\mathcal B}(B^-\to \pi^- ( \rho(1450)^0 \to)K^+ K^-) &=&
(9.46_{-1.65-1.14-0.69-0.38-1.82}^{+1.79+1.16+0.72+0.49+1.82}) \times 10^{-8}, 
\end{eqnarray}
calculated by 
PQCD approach~\cite{Wang:2020nel}.  The latter branching fraction of $B^-\to \pi^- ( \rho(1450)^0 \to)K^+ K^-$
was measured by  LHCb~\cite{LHCb:2019xmb}, Belle~\cite{Belle:2017cxf}, and BABAR~\cite{BaBar:2007itz}. 
Therefore the virtual contribution of $B^-\to \pi^- ( \rho(770)^0 \to)K^+ K^-$ should also 
be considered in experimental analysis.  
%


 Comparing the virtual effect contributions of $\rho$  with
$\omega$ resonance in Table \ref{Pomega}, for instance, $B^-\to \pi^- (\rho^0\to) K^+K^-$ 
and $B^-\to \pi^- (\omega\to) K^+K^-$, we find that the virtual effects are at the same order magnitude, 
even though the decay widths of  $\rho$ and $\omega$ meson are very different, 
 shown in Table~\ref {tab:mass and width}. 
It means that the virtual contributions of these decays are not very sensitive to the 
full decay width of $\rho$ and $\omega$ meson, which is also confirmed by the PQCD approach~\cite{Wang:2020nel}.
 This can be explained by the behavior of the
Breit-Wigner propagators in Eq. (\ref{RBW}) in the kinematics regions of these decays, 
where the imaginary part $\mathrm{i}\, m_{\rho,\, \omega} \Gamma_{\rho,\, \omega}(s)$ 
becomes unimportant when the invariant mass square $s$ is larger than $1 \mathrm{GeV}^2$.

\section{$\it{CP}$ Asymmetry}

\begin{table}[tbhp]
\caption{Direct $\it{CP}$ asymmetry of quasi-two-body decay $B_{(s)} \to P_1 \rho \to P_1 \pi \pi$ 
together with experimental data~\cite{Workman:2022ynf}.}
 \label{PrhoCP}
\begin{center}
\begin{tabular}{cccc}
 \hline \hline
{Modes}     &  Amplitudes & Quasi-two-body results       &  Experiment   \\
\hline
  $B^- \to \pi^- (\rho^0\to)\pi^+ \pi^-$  & $T,C^{\prime}, P, P_A, P_\mathrm{EW}$ &$-0.43\pm0.04$   &$0.009\pm0.019$\\
  
     $B^- \to \pi^0(\rho^-\to)\pi^- \pi^0$ & $T, C, P, P_A, P_\mathrm{EW}$  &$0.15\pm0.02 $ &$0.02\pm0.11$ \\

     $\bar {B}^0 \to \pi^-(\rho^+\to)\pi^+ \pi^0$     &$T, E, P, P_A $         &$-0.43\pm0.03$  &$-0.08\pm0.08$   \\  
                             
  $\bar{B}^0 \to \pi^+(\rho^-\to)\pi^+ \pi^0$           &$T, E, P $         &$0.14\pm0.03$  &$0.13\pm0.06$\\
      
     $\bar {B}^0 \to \pi^0(\rho^0\to)\pi^+ \pi^-$           &$C^{(\prime)}, E, P, P_A,  P_\mathrm{EW}$    &$0.34\pm0.08$  &$-0.27\pm0.24$ \\
                                                           
 \hline 
   $B^- \to K^-(\rho^0\to)\pi^+ \pi^-$             &$T, C^{\prime}, P, P_\mathrm{EW} $~~~         &$0.62\pm0.06$  &$0.37\pm0.10$\\
  
    $B^- \to \bar{K}^0(\rho^-\to)\pi^+ \pi^0$            &$P$         &$0.009\pm 0.000$ &$-0.03\pm0.15$\\ 
                       
  $\bar{B}^0 \to K^-(\rho^+\to)\pi^- \pi^0$            &$T, P$         &$0.59\pm0.01$  &$0.21\pm0.11$\\
     
      $\bar {B}^0 \to \bar{K}^0(\rho^0\to)\pi^+ \pi^-$            &$C^{\prime}, P, P_\mathrm{EW}$         &$-0.085\pm0.059$  &$-0.06\pm 0.09 $\\
                            
  $\bar{B}_s^0 \to K^+(\rho^-\to)\pi^+ \pi^0$           &$ T, P, P_A$         &$0.15\pm0.03$  &...\\
                                                                
  $\bar{B}_s^0 \to K^0(\rho^0\to)\pi^+ \pi^-$       &$C^{\prime}, P, P_C^{\prime}, P_A, P_\mathrm{EW}$         &$-0.40\pm0.14$  &...\\                                
 \hline
   $B^- \to \eta(\rho^-\to)\pi^+ \pi^0$            &$T, C, P, P_C, P_A, P_\mathrm{EW}$~~~        &$-0.11\pm0.02$  &$0.11\pm0.11$\\
     
      $B^- \to \eta^{\prime}(\rho^-\to)\pi^+ \pi^0$   &$T, C, P, P_C, P_A, P_\mathrm{EW}$         &$0.42\pm0.05$  &$0.26\pm0.17$\\
                                 
  $\bar{B}^0 \to \eta(\rho^0\to)\pi^+ \pi^-$  &$C^{(\prime)}, E, P, P_C^{(\prime)}, P_A,  P_\mathrm{EW}$    &$-0.22\pm0.03$  &$-0.23\pm0.03$ \\
   
    $\bar{B}^0 \to \eta^{\prime}(\rho^0\to)\pi^+ \pi^-$   &$C^{(\prime)}, E, P, P_C^{(\prime)}, P_A,  P_\mathrm{EW}$   &$0.083\pm0.078$ &...\\
                                  
  $\bar{B}_s^0 \to \eta(\rho^0\to)\pi^+ \pi^-$          &$C^{\prime}, E, P_C^{\prime}, P_\mathrm{EW}$         &$-0.50\pm0.39$  &...\\
                                                                                      
  $\bar{B}_s^0 \to \eta^{\prime}(\rho^0\to)\pi^+ \pi^-$ &$C^{\prime}, E, P_C^{\prime}, P_\mathrm{EW}$         &$-0.64\pm0.09$  &...\\
                                 
 \hline \hline
\end{tabular}
\end{center}
\end{table}

\begin{table}[tbhp]
\caption{The same as Table \ref{PrhoCP}, but for the quasi-two-body of quasi-two-body decays 
$B_{(s)} \to P_1 K^{*} \to P_1 K \pi$.}
 \label{PKstarCP}
\begin{center}
\begin{tabular}{cccc}
 \hline \hline
{Modes}     &  Amplitudes & Quasi-two-body results       &  Experiment   \\
\hline
$B^- \to \pi^-(\bar{K}^{*0}\to)K^- \pi^+$     &~~~$P, P_A$~~~     &$0.006\pm0.001$    &$-0.04\pm0.09$ \\

   $B^- \to \pi^0(K^{*-}\to)K^- \pi^0$      &~~~$T, C, P, P_A, P_\mathrm{EW}$~~~     &$0.09\pm0.04$    &$-0.39\pm0.21$ \\
                              
 $\bar{B}^0 \to \pi^+(K^{*-}\to)K^- \pi^0$         &~~~$T, P,P_A$~~~     &$-0.21\pm0.04$    &$-0.27\pm0.04$\\
  
  $\bar{B}^0 \to \pi^0(\bar{K}^{*0}\to)K^- \pi^+$       &~~~$C, P, P_A, P_\mathrm{EW}$~~~  &$-0.27\pm0.05$    &$-0.15\pm0.13$\\
                                                          
  $\bar{B}_s^0 \to \pi^0(K^{*0}\to)K^- \pi^+$        &~~~$C, P, P_\mathrm{EW}$~~~     &$-0.29\pm0.06$   &...\\
\hline
 $B^- \to K^-(K^{*0} \to)K^+\pi^-$         &~~~$P, P_A$~~~     &$-0.11\pm0.02$  &$0.12\pm0.10$ \\
  
   $B^- \to K^0(K^{*-}\to)K^- \pi^0$  &~~~$P$~~~     &$-0.19\pm0.18$    &...\\
   
   $\bar{B}^0 \to K^0(\bar{K}^{*0}\to)K^- \pi^+$         &~~~$P$~~~     &$-0.19\pm0.01$    &...\\

    $\bar{B}^0 \to \bar{K}^0(K^{*0}\to)K^- \pi^+$        &~~~$P, P_A$~~~     &$-0.11\pm0.02$    &...\\
 
   $\bar{B}_s^0 \to K^+(K^{*-}\to)K^- \pi^0$       &~~~$T, E, P, P_A$~~~     &$-0.31\pm0.04$    &...\\
                                             

  $\bar{B}_s^0 \to K^0(\bar{K}^{*0}\to)K^- \pi^+$       &~~~$P, P_A$~~~     &$0.002\pm0.001$   &... \\   
                                  
  $\bar{B}_s^0 \to \bar{K}^0(K^{*0}\to)K^+ \pi^-$         &~~~$P$~~~     &$0.009\pm0.000$    &...\\
  \hline
     $B^- \to \eta(K^{*-}\to)K^- \pi^0$            &$T, C, P, P_C, P_A, P_\mathrm{EW}$~~~        &$-0.18\pm0.02$  &$0.02\pm0.06$ \\
     
 $B^- \to \eta^{\prime}(K^{*-}\to)K^- \pi^0$  &$T, C, P, P_C, P_A, P_\mathrm{EW}$~~~        &$-0.47\pm0.09$  &$-0.26\pm0.27$\\
      
 $\bar{B}^0\to \eta(\bar{K}^{*0}\to)K^-\pi^+$         &$C, P, P_C, P_A, P_\mathrm{EW}$~~~    &$0.067\pm0.012$  &$0.19\pm0.05$\\
      
 $\bar{B}^0 \to \eta^{\prime}(\bar{K}^{*0}\to)K^-\pi^+$   & $C, P, P_C^{(\prime)}, P_A, P_\mathrm{EW}$~~~  &$0.062\pm0.052$  &$-0.07\pm0.18$\\
   
      $\bar{B}_s^0\to \eta(K^{*0}\to)K^+\pi^-$            &$C, P, P_C, P_A, P_\mathrm{EW}$~~~        &$0.58\pm0.12$  &...\\ 
      
    $\bar{B}_s^0\to \eta^{\prime}(K^{*0}\to)K^+\pi^-$            &$C, P, P_C, P_A, P_\mathrm{EW}$~~~        &$-0.45\pm0.10$  &...\\                                                                              
 \hline \hline
\end{tabular}
\end{center}
\end{table}

The study of three-body $B$ decays attracts a lot of interests because of its large $\it{CP}$ asymmetry. In this work, we will concentrate on direct $\it{CP}$ violation and will not discuss 
mixing-induced $\it{CP}$ violation. 
We also do not consider  integrated $\it{CP}$ 
asymmetry of the whole phase space of three-body $B$ decays, but we only consider the direct $\it{CP}$ asymmetry
of quasi-two-body decay with one resonant as the intermediate state.
Consequently this $\it{CP}$ asymmetry will arise from the interference between 
tree and penguin amplitudes of the two-body resonances but not from the interference between three-body 
nonresonant and resonant contribution or between different resonant states.

It is well known that direct $\it{CP}$ asymmetries are induced by the interference between difference of strong phases
and different CKM phases of different contributions. 
The strong phase is also the major source of direct $\it{CP}$ violation uncertainty as it is mostly from nonperturbative QCD dynamics. As illustrated in Ref.\cite{Zhou:2016jkv} 
for the two-body $B$ decays, the strong phases 
extracted from experimental data  are sufficient to induce correct 
direct $\it{CP}$ asymmetry, while QCD factorization approaches and soft-collinear effect 
theory usually make wrong predictions or no prediction for the direct $\it{CP}$ asymmetries 
due to absence of nonperturbative strong phases, such as the well-known $K \pi$ puzzle. 
The problem in two-body decays will also exist in three-body decays.
For example, the signs of $\it{CP}$ asymmetry of $B^{-} \rightarrow \pi^{-} K^{+} K^{-}$ 
and $B^{-} \rightarrow K^{-} \pi^{+} \pi^{-}$ decay were conflict with experimental data 
under the QCD factorization approach \cite{Cheng:2013dua}.
The authors need to consider final-state rescattering effect by introducing  
an unknown strong phase $\delta$ to account for the sign flip of $\it{CP}$ asymmetry.
Therefore possible $1/m_b$ power corrections, such as final-state interactions,
or other nonperturbative contributions are necessary to correctly describe $\it{CP}$ asymmetry.
For the quasi-two-body decays under the framework of FAT, the strong phases result in from the 
nonperturbative contribution parameters in Eq.(\ref{parameter}) and the Breit-Wigner formalism 
for resonance in Eq.(\ref{RBW}).  These strong phases have already 
contained all perturbative and nonperturbative QCD effects, for instance, final-state interactions,  
color suppressed contribution, penguin annihilation contribution, and so on. 
They explain well the measured branching ratios, 
and thus are expected to give the right prediction of $\it{CP}$ 
asymmetry parameters in three-body $B$ decays. 

We show the $\it{CP}$ asymmetries of $B_{(s)} \to P_1 \rho \to P_1 \pi \pi$ and 
$B_{(s)} \to P_1 K^* \to P_1 K \pi$ in Tables \ref{PrhoCP} and \ref{PKstarCP}, respectively.
As the theoretical uncertainty from hadronic parameters (form factors and decay constants )
mostly cancel, the major theoretical uncertainties for $\it{CP}$ asymmetry parameters
are from the nonperturbative contribution parameters in Eq.(\ref{parameter}) and weak phases. 
Here we only list the uncertainty arise from the strong phase in Eq.(\ref{parameter}) due to 
small uncertainties of CKM matrix elements. We also list the experimental data in the last column of 
the two tables. Actually only three decay modes with the intermediate process 
$K^- \rho^0$, $\pi^+ K^{*-}$, and $\eta \bar{K}^{*0}$ are well measured in experiments 
with more than $3\sigma$ signal significance. Most $\it{CP}$ asymmetries listed here are predictions 
to be tested by the future experiments. For example,   the $\it{CP}$ asymmetry of 
$B^- \to \pi^+ \rho^0 \to \pi^+ \pi^+ \pi^-$ is predicted quite large 
in this work, which is consistent with the theoretical predictions using 
QCDF~\cite{Cheng:2009cn} and the PQCD approach~\cite{Li:2016tpn}
but not yet measured by the experiments.
We did not list the $\it{CP}$ asymmetries of $B \to P_1 \phi \to P_1 K\bar K$ decays in our table, 
since most of them are from pure penguin diagram contributions, $P_C$, $P_A$,
whose $\it{CP}$ asymmetry is expected to be zero with only one single CKM phase, at leading order approximation. 
We also do not list the $\it{CP}$ asymmetries of $B_{(s)} \to P_1 \rho (\omega) \to P_1 K\bar K$ decays
 contributed by the virtual effects.

 Most of the quasi-two-body decays  are extracted from the Dalitz-plot analysis of three-body ones in experimental analysis. 
For example, $B^- \to K^- \rho^0$ and $B^- \to  \pi^- \bar{K}^{*0}$ decays are extracted from 
$B^- \to K^- \pi^+ \pi^-$ decays. Comparing the $\it{CP}$ asymmetries of quasi-two-body decay 
by integrating out the invariant mass square ($s$) distribution of $\it{CP}$ asymmetry over the 
kinematics region in the Dalitz plot with the corresponding ones of two-body decay with 
$s$ fixed as $m_V^2$ as in Ref~ \cite{Zhou:2016jkv}, we find that the size of 
$\it{CP}$ asymmetries 
of $B_{(s)} \to P_1 \rho \to P_1 \pi \pi$ in Table \ref{PrhoCP} are slightly smaller than that of
the $B_{(s)} \to P_1 \rho $.   The $\it{CP}$ asymmetry of $B_{(s)} \to P_1 K^* \to P_1 K \pi$ in 
Table \ref{PKstarCP} are slightly larger than that of $B_{(s)} \to P_1 K^* $ decay. These very small 
changes of $\it{CP}$ asymmetry between two-body and quasi-two-body decays arise
from the finite decay width of vector resonances.


 \section{Conclusion}\label{sec:4}

We systematically analyze the three-body charmless $B$ meson decays through 
the intermediate resonance, i.e., they proceed via quasi-two-body decays as 
$B_{(s)} \to P_1 V \to  P_1 P_2  P_3$, with the vector resonant state $V$ 
including all ground states $\rho, K^*, \omega, \phi$. 
The first step of  two-body $B$ decays to $P_1 V$ intermediate   states is induced 
by flavor changing weak decays $b \rightarrow \,u\, \, \bar{u} \, d(s)\,$ at leading order 
and $b \rightarrow \,d (s)\, \, q\, \bar{q} \, \, (q=u, d, s)$ at one loop level. 
The second step is that the intermediate vector resonant state $V$, described by 
the Breit-Wigner propagator,  decays into two light 
pseudoscalar mesons  via strong interaction. In order to include all possible perturbative and nonperturbative QCD corrections,
the two-body weak $B$ decays are described by the FAT approach with the decay amplitudes extracted from experimental data. 

We compare results of the branching fractions from FAT approach with the PQCD approach's predictions 
and the ones of QCD factorization, as well as the experimental data. 
For a color suppressed tree diagram and color suppressed or annihilation penguin diagram dominated decay modes, the branching ratios in FAT approach are larger than that of the PQCD approach and the     QCD factorization approach. 
The reason is that 
the decay magnitudes and phases  extracted from experimental data in FAT approach are 
  larger than  that in the other two approaches due to
the shortage of   nonperturbative contribution and $1/m_b$ power corrections.
We have also considered the virtual effects from $\rho$ and $\omega$   resonance tail, which are 
usually ignored by the experimental analysis.   These virtual effects from ground state 
$\rho(770)^0$ are about $1\sim2$ order magnitude smaller than the dominated contribution of 
$B_{(s)} \to \pi, K ( \rho \to) \pi \pi$ decays. However, these virtual effects contributions are at the same order as the higher resonance contributions, such as $\rho (1450)$. We also find that the virtual contributions of 
these decays are not very sensitive to the decay widths of $\rho$ and $\omega$.
The branching ratios of $B_{(s)} \to P_1 (\rho \to ) \pi \pi $, $B_{(s)} \to P_1 (K^* \to ) K \pi $
are consistent with experimental data, and the others are predictions waiting for future 
experiments to test.

For $\it{CP}$ asymmetry, theoretical predictions of three-body decays are mostly based on quasi-two-body 
decays via intermediate resonance, where the strong phases extracted from experimental 
data in FAT approach are sufficient to induce correct direct $\it{CP}$ asymmetries.
In this work, we utilize the same strong phases of two-body decays together with 
the phases from Breit-Wigner formalism to predict the direct $\it{CP}$ asymmetries 
of quasi-two-body decays, $B_{(s)} \to P_1 ( \rho \to) \pi \pi$ and 
$B_{(s)} \to P_1 ( K^* \to) K \pi$. 
Our results agree well with the existing experimental measurements. But many of the $\it{CP}$
 asymmetry predictions are waiting for future experiments.


\section*{Acknowledgments}
S.-H. Z. thanks Theoretical Physics Division of Institute of High Energy Physics, CAS for hospitality 
while this work was completed. 
The work is supported by the National Natural Science Foundation of China
under Grants No. 11765012, No. 12075126, No. 12275277, No. 12070131001,
and No. 12105148,  as well as the National Key Research and Development Program of 
China under Contract No. 2020YFA0406400.

\bibliographystyle{bibstyle}
\bibliography{refs}

\end{document}